\documentclass[letterpaper]{article} 
\usepackage{aaai2026}  
\usepackage{times}  
\usepackage{helvet}  
\usepackage{courier}  
\usepackage[hyphens]{url}  
\usepackage{graphicx} 
\usepackage{natbib}  
\usepackage{caption} 
\usepackage{algorithm}
\usepackage{algorithmic}
\usepackage{multirow}

\usepackage{xcolor}
\newcommand{\answerYes}[1]{\textcolor{teal}{#1}} 
 
\newcommand{\answerNA}[1]{\textcolor{gray}{#1}}

\usepackage{newfloat}
\usepackage{listings}
\usepackage{subcaption}

\newcommand{\rev}[1]{\textcolor{black}{#1}}
\newcommand{\revii}[1]{\textcolor{black}{#1}}

\DeclareCaptionStyle{ruled}{labelfont=normalfont,labelsep=colon,strut=off} 
\floatstyle{ruled}
\newfloat{listing}{tb}{lst}{}
\floatname{listing}{Listing}
\usepackage{booktabs}

\usepackage{amsmath}
\usepackage{tcolorbox}
\usepackage{enumitem}
\tcbuselibrary{listings}
\usepackage{tcolorbox}
\usepackage{afterpage}
\usepackage{cellspace}
\usepackage{pbox}
\usepackage{makecell}

\newtcblisting{promptbox}[1]{%
  title=#1,
  listing only,
  colback=white!95!gray,
  colframe=gray,
  width=\columnwidth,
  arc=0mm,
  boxrule=0.5mm,
  colbacktitle=white!95!gray,
  coltitle=black,
  fonttitle=\bfseries,
  left=6pt,
  right=6pt,
  top=6pt,
  bottom=6pt,
  listing options={basicstyle=\small\ttfamily, breaklines=true}
}

\title{Framing Unionization on Facebook: Communication Around Representation Elections in the United States}
\author{
    Arianna Pera\textsuperscript{\rm 1},
    Veronica Jude\textsuperscript{\rm 2},
    Luca Maria Aiello\textsuperscript{\rm 1,3},
    Ceren Budak\textsuperscript{\rm 2}\\
}
\affiliations{
    \textsuperscript{\rm 1}IT University of Copenhagen\\
    \textsuperscript{\rm 2}University of Michigan\\
    \textsuperscript{\rm 3}Pioneer Centre for AI\\

    arpe@itu.dk, vjude@umich.edu, luai@itu.dk, cbudak@umich.edu

}

\begin{document}

\maketitle

\begin{abstract}
Digital media have become central to how labor unions communicate, organize, and sustain collective action. Yet little is known about how unions’ online discourse relates to concrete outcomes such as representation elections. This study addresses the gap by combining National Labor Relations Board (NLRB) election data with 158k Facebook posts published by U.S. labor unions between 2015 and 2024. We focused on five discourse frames widely recognized in labor and social movement communication research: diagnostic (identifying problems), prognostic (proposing solutions), motivational (mobilizing action), community (emphasizing solidarity), and engagement (promoting \rev{social media} interaction). Using a fine-tuned RoBERTa classifier, we systematically annotated unions’ posts and analyzed patterns of frame usage around election events. 
Our findings showed that diagnostic and community frames dominated union communication overall, but that frame usage varied substantially across organizations. 
\rev{Greater use of diagnostic, prognostic, and community frames prior to an election was associated with higher odds of a successful outcome. After elections, framing patterns diverged depending on results: after wins, the use of prognostic and motivational frames decreased, whereas after losses, the use of prognostic and engagement frames increased.}
By examining variation in message-level framing, the study highlights how communication strategies \rev{correlate with organizational success}, contributing open tools and data, and complementing prior research in understanding digital communication of unions and social movements.
\end{abstract}

\section{Introduction}
Digital media has transformed the landscape of collective action. 
Civil Society Organizations (CSOs)\rev{, interest groups,} and social movements increasingly rely on online platforms to mobilize participants, coordinate activities, and frame issues to reach broader audiences. 
Scholars have highlighted the promises of these digital technologies: lowering barriers to participation, enabling new forms of connective action, and amplifying marginalized voices. At the same time, they have cautioned about their limitations, including resource constraints, fragmented attention, and uncertain effects on long-term organizational capacity~\cite{bennett2012logic, earl2011digitally, tufekci2017twitter}. 

Labor unions provide a distinctive case within \rev{this} ecosystem. 
As one of the most institutionalized forms of collective organization, they have experienced declining membership and political influence in the United States since the mid-twentieth century. This led scholars to ask whether their revitalization requires not only traditional collective bargaining but also new forms of communication and public engagement~\citep{milkman2013back,hurd1998union,voss2000breaking}. 
\rev{Given today’s fragmented workforce with precarious employment and immigrant labor~\citep{milkman2020immigrant}, social media platforms offer a strategic opening for unions.
Accordingly, unions have adopted digital media as a communication infrastructure to mobilize members, reach new audience, and maintain organizational relevance~\cite{lee1997labour}}.

Among digital platforms, Facebook has emerged as particularly important~\cite{carneiro2022digital, ford2022digital}. 
Like other social media, it supports both outward and inward communication, and attracts users whose demographic profiles resemble many unionized workforces. These features make it strategically relevant for labor mobilization~\cite{blanc2022digitized}.
\rev{In this study, we focus on Facebook for both substantive and practical reasons. Among the 77 unions affiliated with the AFL-CIO\footnote{AFL-CIO affiliated unions. \url{https://aflcio.org/about-us/our-unions-and-allies/our-affiliated-unions}. Accessed March 3, 2025.} (the largest federation of unions in the United States), 63 link to Facebook, 66 to Twitter, and 43 to YouTube. Although Twitter is slightly more prevalent, its data has become increasingly restricted~\cite{murtfeldt2024rip}. Facebook therefore offers both broad adoption and comparatively accessible data, making it well suited for empirical analysis.}
\rev{Despite its potential, much remains unknown about how unions use the platform effectively: there is }uncertainty about which engagement strategies are most effective~\cite{greene2003possibilities}, and about how union communication on the platform relates to concrete unionization outcomes at scale.

In this study, we combined National Labor Relations Board (NLRB) election data with a large corpus of Facebook posts by labor unions in the United States from 2015 to 2024 to examine the \rev{correlation} between online communication and \rev{representation} election outcomes. 
\rev{Representation elections are administered by the NLRB and allow eligible employees to vote on whether a union should be certified as their exclusive bargaining representative. Such elections occur regularly across industries, initiated by employee petitions.}
Using natural language processing and computational methods, we analyzed how unions framed their digital communication around representation elections and whether these patterns varied by electoral outcomes\footnote{Data and code are available at 
\\\url{https://github.com/ariannap13/laborunions}}. Specifically, we addressed three research questions:

\vspace{2pt} \noindent \textbf{RQ1.} \emph{To what extent do labor unions use discourse frames in their Facebook communication?}

\vspace{2pt} \noindent \textbf{RQ2.} 
\emph{
\rev{How is the use of different discourse frames prior to an election associated with the likelihood of electoral success?}}

\vspace{2pt} \noindent \textbf{RQ3.} \emph{How do patterns of frame usage change after election events, and do these dynamics vary between wins and losses?}
\vspace{2pt}

\noindent \rev{While our primary analysis focused on discourse frames, prior research on communication suggests that the emotional content can shape mobilization~\cite{feldman2016using} and that the topics discussed may influence audience response~\cite{wang2021hashtag}. 
We therefore complemented our frame-based analysis with emotion detection and topic modeling to provide a more nuanced view of union communication.}

Our contributions are threefold: (i) we provide guidelines for annotating discourse frames in unions' social media text, (ii) we release a fine-tuned, ready-to-use classification model for detecting frames in unions' communication; and (iii) we advance research at the intersection of framing and labor movement, offering insights into the relationship between digital communication and unionization dynamics. \rev{Although} much of labor movement scholarship treats unions at the collective level~\cite{snow1988ideology, benford2000framing, kelly1998rethinking, voss2000breaking}, we show how digital data enables the observation of framing strategies at the level of individual messages and connect them to concrete organizational outcomes. This micro-comparative approach \rev{fits well with} the heterogeneity of communication practices across unions, and allows us to \rev{study how framing varies during periods of organizational change}.

\rev{Building on this perspective, our analysis of union communication on Facebook revealed several key patterns}. We found that unions predominantly emphasized problem identification and community-building, though frame use varied across organizations. The mobilization potential, embodied by a discourse centered on call-to-action messages, remained largely untapped. 
\rev{This pattern echoes prior research showing that unions and CSOs tend to prioritize problem framing and information dissemination on social media, while making more limited use of call-to-action messages~\cite{lovejoy2012information, uba2021political, carneiro2022digital}.}
Unions representing broad, multi-industry workforces (\emph{industrial} unions) tended to emphasize frames for grievances and solutions, whereas unions representing specialized trades (\emph{craft} unions) relied less on these frames.
\rev{By fitting a Bayesian logistic regression model to predict representation election outcomes, we found that a greater focus on problem identification, solution-oriented messaging, and community building prior to the election is associated with higher odds of winning. In cases that were won,} solution-oriented and call-to-action frames generally declined after the election. In cases that \rev{were} lost, by contrast, \rev{solution-oriented and interaction-focused frames increased after the election. Overall, these results indicate that election outcomes are associated with distinct patterns of communicative emphasis.}
\rev{These results are based on observational data and do not allow us to make causal claims about the effect of frames on electoral success.}
\rev{The association} between communication strategies and organizational outcomes, as well as their adjustment after victory, is also reflected in prior work on union revitalization~\cite{turner2005transformation} and theories of organizational life cycles of social movements~\cite{christiansen2009four, tarrow2022power}.

Our findings provide actionable insights for labor unions, CSOs, \rev{interest groups, }and social movements on their digital communication strategies. 
In particular, \rev{successful election outcomes are more frequently associated with social media use emphasizing issue discussion, the proposal of solutions, and solidarity rather than engagement prompts (e.g., encouraging likes or shares) or mobilization cues.}
Beyond these practical implications, our results contribute to theoretical debates on framing and collective action by showing how message-level strategies vary across organizations and electoral contexts. By shifting focus from collective-level analyses of movements to the micro-dynamics of discourse on digital platforms, this study \rev{offers a lens for examining how framing may be involved in processes of adaptation during periods of organizational change}.

\section{Theoretical Framework: Discourse Frames}
Labor unions, like other CSOs\rev{, interest groups,} and social movements, rely on communication to achieve core organizational goals \rev{and lobbying}. Nowadays, much of this activity occurs in digital spaces, making it critical to understand the communicative strategies unions employ online.

\rev{
Framing is central to communication studies, but its meaning varies across traditions. In rhetorical approaches, frames selectively emphasize certain considerations to shape audience interpretations~\citep{gamson1989media, tversky1981framing, druckman2001implications, kinder1998communication}. Social movement research treats frames more broadly as recurring points of emphasis that structure problems, propose solutions, and motivate action~\citep{snow1988ideology, benford2000framing}, highlighting meaning-making and collective behavior rather than persuasion. Interest group research similarly sees framing as strategic and instrumental~\citep{baumgartner2009lobbying}, but emphasizes deployment under specific political or organizational conditions rather than exhaustive typologies. Building on these perspectives, we adopt a functional view of frames in union social media, analyzing a selective subset that reflects their role in organizing meaning and facilitating collective action. Although not exhaustive, these categories provide a theoretically grounded and analytically appropriate framework for studying labor union communication online.}

The theory of collective action frames~\citep{snow1988ideology, benford2000framing} offers a central lens for analyzing frames and functions of unions' social media communication: \textbf{diagnostic} frames, which highlight problems and assign responsibility; \textbf{prognostic} frames, which propose solutions; and \textbf{motivational} frames, which seek to mobilize audiences. These three core frames also parallel the conditions for union revitalization outlined by~\citet{kelly1998rethinking}: workers must recognize grievances (diagnostic), perceive them as actionable (prognostic), and see unions as legitimate vehicles for change (motivational).

\rev{To capture the specificity of union communication online, we draw on literature on organizational and movement communication on social media~\citep[e.g.,][]{lovejoy2012information, carneiro2022digital, muntinga2011introducing}, which identifies a comprehensive range of frame types. Guided by iterative analysis of the data, we focus on those categories that are both theoretically meaningful and empirically prevalent. Specifically, we examine two frame types. \textbf{Community} frames emphasize shared identity and solidarity without urging action~\citep{carneiro2022digital, lovejoy2012information}, while \textbf{engagement} frames promote platform-specific interaction (e.g., liking, sharing), reflecting the affordances of social media~\citep{muntinga2011introducing}.}

In summary, our analysis operationalized five discourse frames:
\begin{itemize}[noitemsep]
\item \textbf{Diagnostic}: identifying problems and attributing responsibility.
\item \textbf{Prognostic}: advancing solutions and strategies for change.
\item \textbf{Motivational}: encouraging participation through calls to action \rev{relevant to union goals (e.g., urging followers to join a union or attend protests)}.
\item \textbf{Community}: fostering solidarity and collective identity\rev{, without prompting a specific action}.
\item \textbf{Engagement}: \rev{leveraging social media affordances (likes, shares, comments) to sustain visibility and algorithmic reach, rather than to mobilize action or express collective identity.}
\end{itemize}

This framework maps prior research \rev{on framing and union communication onto an operationalization of key framing functions observable in social media discourse.}

\section{Methodological Framework}

\subsection{Data Curation} 
Our focus is on the central stage of the unionization process: representation elections and their outcomes (win or loss).
We collected data on election results from publicly available records of the NLRB~\cite{NLRB}, covering all cases for which a representation election (\emph{RC} cases in the NLRB data) was petitioned and held between January 2015 and December 2024 for labor unions in the United States. From the available variables, we focused on the numeric identifier of each election case, the election date, the names of participating unions, and the union that won the election.

The dataset comprises 11,416 election cases, of which 10,975 involve a single union, 417 involve two unions, and 24 involve three unions competing in the same election.
\rev{On average, there were 20 elections per week between 2015 and 2024. Once elected, a union remained the representative for a given set of workers for an average of 237 days.}
Since much of the information was entered manually in the NLRB data, union names exhibit considerable variation in formatting and granularity. To facilitate analysis of unions’ Facebook (FB) activity, we standardized these names at the national or international level. For example, ``Communications Workers of America Local 7250'' was mapped to ``Communications Workers of America.''
To identify a set of national and international unions, we began with the official list of unions affiliated with the AFL-CIO, consolidating subunits under their parent organizations (e.g., \emph{TNG-CWA} under \emph{CWA}) using regular expressions and rule-based assignments. This process yielded 47 unions. 
For unions not included in the AFL-CIO list, we identified common patterns and acronyms using a combination of regular expressions (e.g., extracting acronyms in parentheses) and manual grouping\rev{. This resulted in additional 510 potential national and international unions, including a substantial amount of noise from local chapters which we could not easily reconduct to parent unions}.

To link these organizations with their digital presence, we queried the Bing Search API\footnote{Microsoft Bing Search API. \url{https://api.bing.microsoft.com/v7.0/search}. Accessed March 3, 2025} with each union name alongside the string ``union Facebook'' to identify official FB pages. We parsed the results using regular expressions and manually verified their accuracy. This resulted in 66 unions with official FB \rev{pages}, of which 40 provided \rev{downloadable} data through the Meta Research API Content Library~\cite{META}\footnote{\rev{According to the API terms, data are only downloadable for pages with at least 15,000 followers or likes.}}. 
\rev{While the resulting number of unions included in the dataset may appear limited, these unions collectively represent approximately 46\% of all union members in the United States in the 2024 fiscal year\footnote{\rev{Data from \url{https://www.dol.gov/agencies/olms/public-disclosure-room}}}.}
Table~\ref{tab:union-acronyms} lists the unions involved in the study, \rev{along} with their acronyms and type of structure. In total, we collected 158,238 posts (157,883 of which had text available) published between January 2015 and December 2024, matching the time span of the election data.

Of the 11,416 NLRB election cases, we identified 2,607 instances where unions with available FB data lost and 6,445 where they won. 

\paragraph{\rev{Control Factors}}
\rev{We collected data on union characteristics and social media activity to account for factors that may influence election outcomes.}

\rev{Traditional union-level variables were drawn from public records published by the U.S. Department of Labor\footnote{\url{https://www.dol.gov/agencies/olms/public-disclosure-room}} and by the Center for Union facts\footnote{\url{https://unionfacts.com/find-your-union/}}. These include the number of members, assets, disbursements, liabilities, and primary industry (e.g., construction, educational services).
From the NLRB election data, we recorded the state in which each election occurred to capture broader state-level institutional and political context. Separately, we coded whether the election took place in a \emph{right-to-work} jurisdiction, capturing a specific legal dimension of state labor regimes.
Finally, following a common distinction in labor movement scholarship~\cite{herrigel1993identity,dubofsky2000we}, we annotated unions for their organizational type (craft vs. industrial) through manual review of their official websites.}

\rev{For social media activity, we drew on the collected Facebook data, recording follower counts and the total number of posts over the analyzed period. From the latter, we computed the average number of posts per week. To reduce multicollinearity, we evaluated Variance Inflation Factors (VIFs) and excluded variables with $\text{VIF} \geq 5$, resulting in the removal of union employees and Facebook followers from the final set of controls.}

\subsection{Discourse Frames Classifier}
We trained a classifier to detect five discourse frames in union Facebook posts: \emph{diagnostic}, \emph{prognostic}, \emph{motivational}, \emph{community}, and \emph{engagement}. Aiming at capturing the nuanced usage of such frames in online communication, we set the task to be multi-label.
Following common practices in Natural Language Processing (NLP), we first fine-tuned a RoBERTa-base model~\cite{liu2019roberta} on the text of the 157,883 posts to better capture typical linguistic patterns of labor union communication on Facebook. After collecting annotations for the presence of discourse frames\footnote{We also annotated posts for an additional \emph{political endorsement} label to capture explicit support for political actors. Given its low prevalence, it was not included in the main analysis.}, we further fine-tuned this model to perform the multi-label classification task. \rev{We also included a RoBERTa-base model directly fine-tuned on the frame detection task, and a BERT-base model as additional classifiers for comparison.}
Details on model configuration and computational resources are provided in Appendix \ref{app:modeling_resources}.

\paragraph{Frames Annotation}
We conducted a manual annotation campaign with two of the authors as coders, considering a multi-label setting for the presence of discourse frames. Starting with an initial set of 100 posts and taking inspiration from prior work on measuring collective action frames in text~\cite{fernandez2023digital, mendelsohn2024framing}, we iteratively defined a codebook for the identification of discourse frames of interest. The codebook is reported in Appendix~\ref{app:annotation_codebook}.

Using this codebook, the coders annotated a new\rev{, random} sample of 200 posts, achieving a Cohen’s $\kappa$ inter-annotator agreement of \rev{0.81 averaged across labels\footnote{\rev{Per-label agreement: 0.88 (diagnostic), 0.86 (prognostic), 0.86 (motivational), 0.62 (community), and 0.85 (engagement)}}}.
Given this level of agreement, the coders then independently annotated disjoint sets of additional randomly sampled posts. In total, this resulted in 1,151 new posts, each annotated by a single coder. 
\rev{The annotated data is temporally well distributed: posts from 2016 to 2024 each account for between 5\% and 14\% of the sample, while 2015 contributes just over 2\%.}

\paragraph{Multi-label Classification}
We randomly split the annotated dataset into training (80\%) and test (20\%) sets. The RoBERTa model, pre-adapted to Facebook language, was then fine-tuned on the training set for the multi-label classification task. After evaluating its performance on the test set, we applied the final classifier to the entire corpus to generate frame predictions for all posts.

\subsection{\rev{Supplementary Linguistic Measures}}
\rev{In addition to coding frames, we also computed two supplementary linguistic measures to characterize posts in more detail. First, we generated an \textbf{emotional profile} for each post based on Ekman’s six basic emotions~\cite{ekman1992argument} plus a neutral category, using a pre-trained DistilRoBERTa classifier\footnote{\url{j-hartmann/emotion-english-distilroberta-base} on HuggingFace} to produce a probability vector over the seven emotional categories. Second, we applied \textbf{topic modeling} to posts using BERTopic~\cite{grootendorst2022bertopic} with DistilRoBERTa embeddings~\cite{sanh2019distilbert}.}

\subsection{Patterns of Frames Usage} \label{subsec:frames_usage}
We examined how unions’ usage of discourse frames \rev{prior to elections can contribute to the prediction of elections' outcomes}, and how \rev{frame usage changed after the elections}.

\rev{To account for sparsity in posting frequency, we aggregated daily frame usage over short, overlapping windows of $n_{days}=5$}. Given an active union on Facebook, for each day and each frame, we calculated (i) the number of posts containing that frame and (ii) the proportion of such posts relative to all posts on that day. To focus on activity around elections, for each election case, we defined two non-overlapping \rev{aggregated measures}:
\begin{itemize}[noitemsep]
    \item \rev{$\text{\textbf{pre}}_{fi}$: mean proportion of posts expressing frame $f$ from 7 to 3 days before election $i$}.
    \item \rev{$\text{\textbf{post}}_{fi}$: mean proportion of posts expressing frame $f$ from 3 to 7 days after election $i$}.
\end{itemize}
\noindent These \rev{aggregated measures} exclude the election day itself---avoiding overlap caused by the \rev{moving} aggregation---while capturing both the immediate lead-up and aftermath. \rev{The aggregation does not introduce temporal smoothing, as analyses rely on fixed pre- and post-election summaries.}

\paragraph{\rev{Predicting Election Outcome}}
\rev{To examine whether the use of specific frames was associated with electoral success, we fitted a Bayesian logistic regression model to predict the electoral outcome (win vs. loss). In this framework, regression coefficients are estimated as posterior distributions, providing a full distribution of plausible values for each effect. Weakly informative priors were automatically assigned to all coefficients using the \texttt{bambi} Python package. The model was specified as follows:}
\begin{equation*}
\begin{aligned}
\text{logit}\big(\Pr(\text{election\_win}_i = 1)\big)
= \beta_0 + \sum_{f \in F} \beta_f \text{pre}_{fi} \\
+ \sum_{e \in E} \beta_e \text{emo}_{ei} 
+ \sum_{c \in C} \beta_c X_{ci}
+ u_{\text{union[i]}} + u_{\text{election-state[i]}}
\end{aligned}
\end{equation*}

\rev{where $election\_win_{i}$ is the binary indicator of the outcome for election $i$, and $\beta_0$ is the intercept. 
Predictors fall into four groups: framing variables, other linguistic characteristics, traditional union-level controls, and online presence controls.}

\rev{Framing variables are captured by $\text{pre}_{fi}$, the proportion of posts prior to election $i$ using frame $f$ (diagnostic, prognostic, motivational, community, or engagement), with coefficients $\beta_f$.
Other linguistic characteristics are represented by $\text{emo}_{ei}$, the mean predicted probability that posts prior to election $i$ expressed emotion $e$ (anger, disgust, fear, joy, sadness, surprise, or neutral), with coefficients $\beta_e$. We do not include topic-level predictors in this analysis because most topics are union-specific (see Results).}

\rev{Control variables $X_{ci}$ with coefficients $\beta_c$ include traditional union-level characteristics and online presence measures. These divided into traditional union level controls and online presence controls. Union-level controls comprise union size, assets, disbursements, liabilities, primary industry, organizational structure (craft vs. industrial), and the presence of a right-to-work law in the election state.
Online presence is measured by the average number of weekly posts. All numerical controls are standardized.
We also included normally distributed random intercepts for unions, $u_{\text{union[i]}} \sim \mathcal{N}(0, \sigma_u^2)$, and election states, $u_{\text{election-state[i]}} \sim \mathcal{N}(0, \sigma_u^2)$, to account for unobserved heterogeneity.}

\paragraph{\rev{Detrended Frame Usage: Changes Pre- vs. Post-Election}}
\rev{To examine how unions adjusted their communication around elections, we computed detrended frame usage to remove long-term trends. For each union, baseline periods of 18 consecutive days were identified during January 2015--December 2024 when the union was not involved in a specific election. A pseudo-event date was assigned at the midpoint of each baseline period, and pre- and post-event aggregation was applied like for real elections.}

We then computed the detrended usage of \rev{frame $f$} as
\begin{align*}
{\text{pre}_{fi}}^{D} &= \text{pre}_{fi}-{\text{pre}_{fi}}^{baseline} \\
{\text{post}_{fi}}^{D} &= \text{post}_{fi}-{\text{post}_{fi}}^{baseline}
\end{align*}

where $\text{pre}_{fi}$ and $\text{post}_{fi}$ denote \rev{usage of frame $f$} before and after the election \rev{$i$}, and ${\text{pre}_{fi}}^{baseline}$ and ${\text{post}_{fi}}^{baseline}$ are the corresponding baseline values. Values of ${\text{pre}_{fi}}^{D}$ or ${\text{post}_{fi}}^{D}$ close to 0 indicate frame usage similar to baseline periods, positive values indicate higher-than-baseline usage, and negative values indicate lower-than-baseline usage. 

\rev{The change from pre- to post-election was computed as}
\[O = {\text{pre}_{fi}}^{D}-{\text{post}_{fi}}^{D}\]

\rev{Union-case instances were classified as \textbf{decrease}, \textbf{stable}, or \textbf{increase} using the 25th and 75th percentiles of the $O$ distribution:}
\[
O =
\begin{cases}
\textbf{decrease}, & O \leq \text{percentile}_{25} \\
\textbf{stable}, & \text{percentile}_{25} < O \leq \text{percentile}_{75} \\
\textbf{increase}, & O > \text{percentile}_{75}
\end{cases}
\]

This procedure provided a consistent way to categorize changes in frame usage, enabling systematic comparisons between won and lost election cases and across frames.

\paragraph{Robustness Checks}
To ensure that results generalize well beyond specific design choices, we conducted several robustness checks: 
(i) we varied the \rev{moving} aggregation window size ($n_{days}=5$ vs $n_{days}=3$) (see Appendix \ref{app:window_size_rolling}); 
(ii) we balanced the number of won and lost cases within each labor union, repeating analyses across multiple random seeds (see Appendix \ref{app:imbalance_n_cases}); and
(iii) \rev{we down-sampled representation elections from unions with exceptionally high electoral activity, capping the number of cases per union at the 90th percentile of the distribution, to reduce the influence of a small number of highly active unions. Analyses were repeated across multiple random seeds} (see Appendix~\ref{app:imbalance_n_cases}). 

\section{Results}

\subsection{Classification of Discourse Frames}
We applied our fine-tuned RoBERTa model to Facebook posts to perform multi-label classification of discourse frames. Table~\ref{tab:classification-report} reports performance in terms of precision, recall, and F1 scores for the test set. Scores are comparable to similar studies~\cite{fernandez2023digital, mendelsohn2024framing}, with F1 values ranging from 0.68 to 0.78 across frames and a macro F1 of 0.73. 
\rev{To evaluate the performance of this model, we compared it against two alternative classifiers, and found that it performs on par with or better than both (see Appendix~\ref{app:alt_classifier}).}
\setlength{\tabcolsep}{5pt} 
\begin{table}[ht]
\centering
\begin{tabular}{@{}lcccc@{}}
\toprule
\textbf{Frame} & \textbf{Precision} & \textbf{Recall} & \textbf{F1-score} & \textbf{Support} \\
\midrule
Diagnostic            & 0.72 & 0.85 & 0.78 & 84 \\
Prognostic            & 0.67 & 0.69 & 0.68 & 58 \\
Motivational          & 0.57 & 0.83 & 0.68 & 29 \\
Community             & 0.75 & 0.81 & 0.78 & 81 \\
Engagement            & 0.71 & 0.78 & 0.75 & 51 \\
\midrule
Micro avg             & 0.70 & 0.80 & 0.74 & 303 \\
Macro avg             & 0.68 & 0.79 & 0.73 & 303 \\
Weighted avg          & 0.70 & 0.80 & 0.74 & 303 \\
\bottomrule
\end{tabular}
\caption{Performance of the multi-label discourse frames classifier. \rev{\textbf{Support} indicates the number of examples in each class or overall.}}
\label{tab:classification-report}
\end{table}

\noindent We then applied the classifier to the corpus of 157,883 textual posts, identifying 140,199 as expressing at least one of the frames of interest. 

\subsection{Usage of Discourse Frames}
Aiming at defining a baseline distribution of frame usage across unions, we first looked at the distributions of posts by union. We identified skewness (minimum = 779, 25th percentile = 1,921, 75th percentile = 5,442, maximum = 10,452, full distribution in Figure \ref{fig:n_post_union}) and thus implemented a bootstrap-like down-sampling procedure. Over-represented unions were down-sampled to the minimum number of posts in the distribution, and the procedure was repeated with five different random seeds. For each frame, we computed the median proportion of posts expressing that frame across seeds.
Overall, the median baseline distribution of frame usage is as follows: diagnostic in 40.7\% of posts, community in 39.4\%, prognostic in 28.3\%, engagement in 23.2\%, and motivational in 15.4\%. 

\begin{figure}[t!]
    \centering   
    \includegraphics[width=0.475\textwidth]{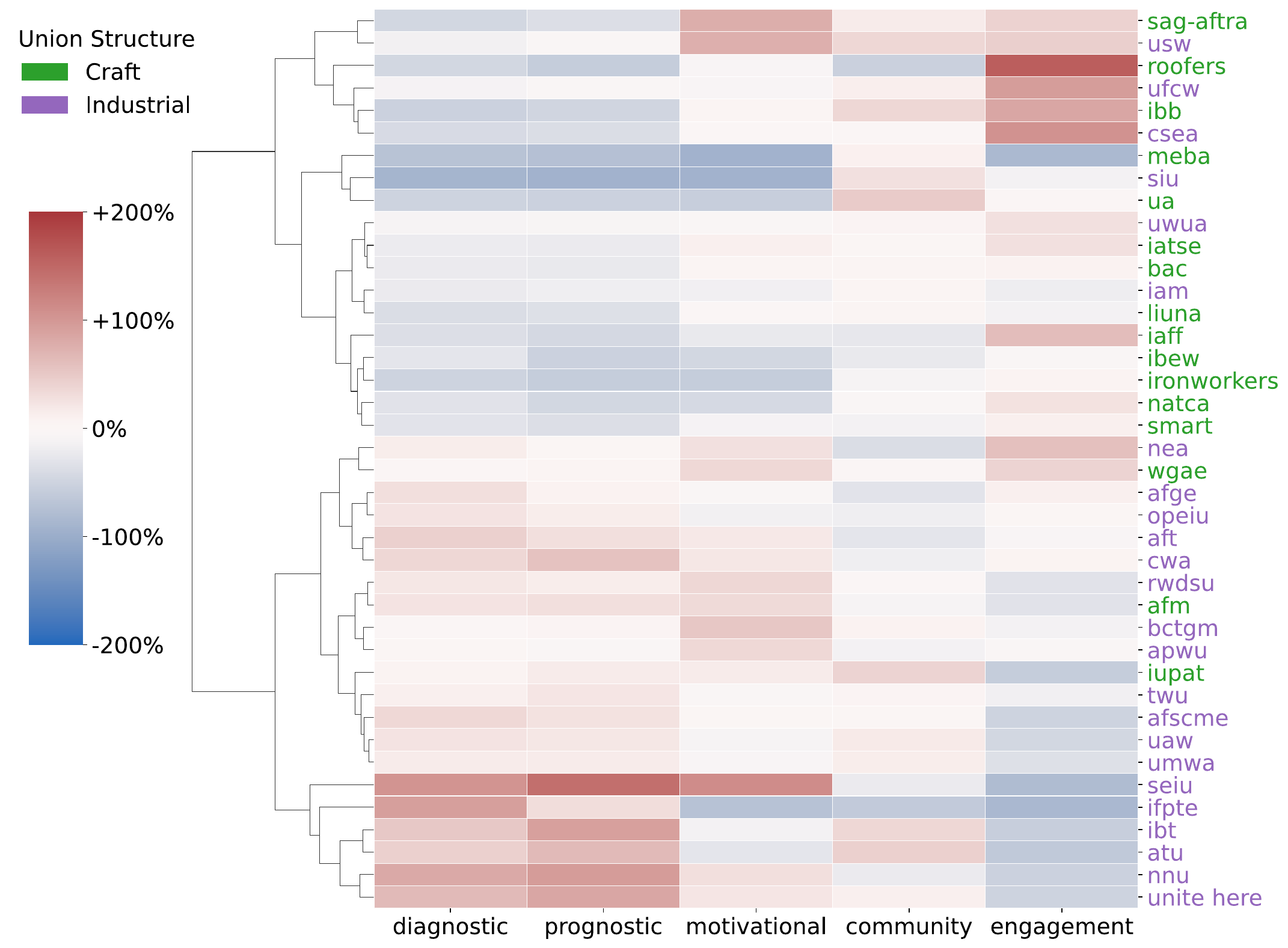}
    \caption{Relative usage of discourse frames by labor unions (percentage difference from baseline). Red indicates above-baseline usage, blue below. The dendrogram clusters unions by framing patterns across the five frames, and acronyms are colored by organizational structure.}
    \label{fig:frames_by-union}
\end{figure}

Figure~\ref{fig:frames_by-union} shows how individual unions diverge from the overall median baseline in their use of different frames.
For each union, we calculated the share of posts expressing a given frame relative to its total output, and then expressed this value as a percentage difference over the median baseline.
A value of +100\% indicates that the union used that frame twice as often as the median baseline, 0\% indicates equal usage, and –50\% indicates that the union used the frame half as often. 

Many unions behaved in line with the baseline; nonetheless, a clear divide emerged. Unions clustered into two broad groups based on whether their use of diagnostic and prognostic frames was higher or lower than the baseline. In particular, industrial unions tended to employ diagnostic and prognostic frames more frequently than the baseline. Examples include the Service Employees International Union (SEIU), the International Federation of Professional and Technical Engineers (IFPTE), the International Brotherhood of Teamsters (IBT), the Amalgamated Transit Union (ATU), National Nurses United (NNU), and Unite Here (UNITE HERE), all located toward the bottom of the dendrogram (union acronyms are expanded in Table~\ref{tab:union-acronyms}). By contrast, lower-than-baseline usage of diagnostic and prognostic frames was more characteristic of craft unions. This distinction might reflect different organizational orientations: industrial unions, with their broad membership bases, are more often engaged in large-scale mobilization and bargaining arenas, where diagnostic frames (grievance articulation) and prognostic frames (solution proposals) play a central role~\cite{voss2000breaking}.
Variation also emerged in the use of the engagement frame. Industrial unions, in particular, often used engagement frames less than the baseline (cf. bottom of the plot), whereas higher-than-baseline engagement showed no consistent relationship to union structure (cf. top of the plot). Notably, the United Union of Roofers, Waterproofers and Allied Workers (\emph{Roofers}) employed the engagement frame nearly two and a half times more often than the baseline.

Taken together, these results suggest that industrial unions tend to differentiate themselves from craft unions by emphasizing grievance and solution-oriented frames while downplaying interactive engagement, consistent with their larger-scale bargaining and mobilization needs. Craft unions, on the other hand, appear less reliant on diagnostic and prognostic repertoires, and in some cases lean more heavily on engagement, indicating a different balance in communication strategies.

\begin{figure}[t!]
    \centering   
    \includegraphics[width=0.47\textwidth]{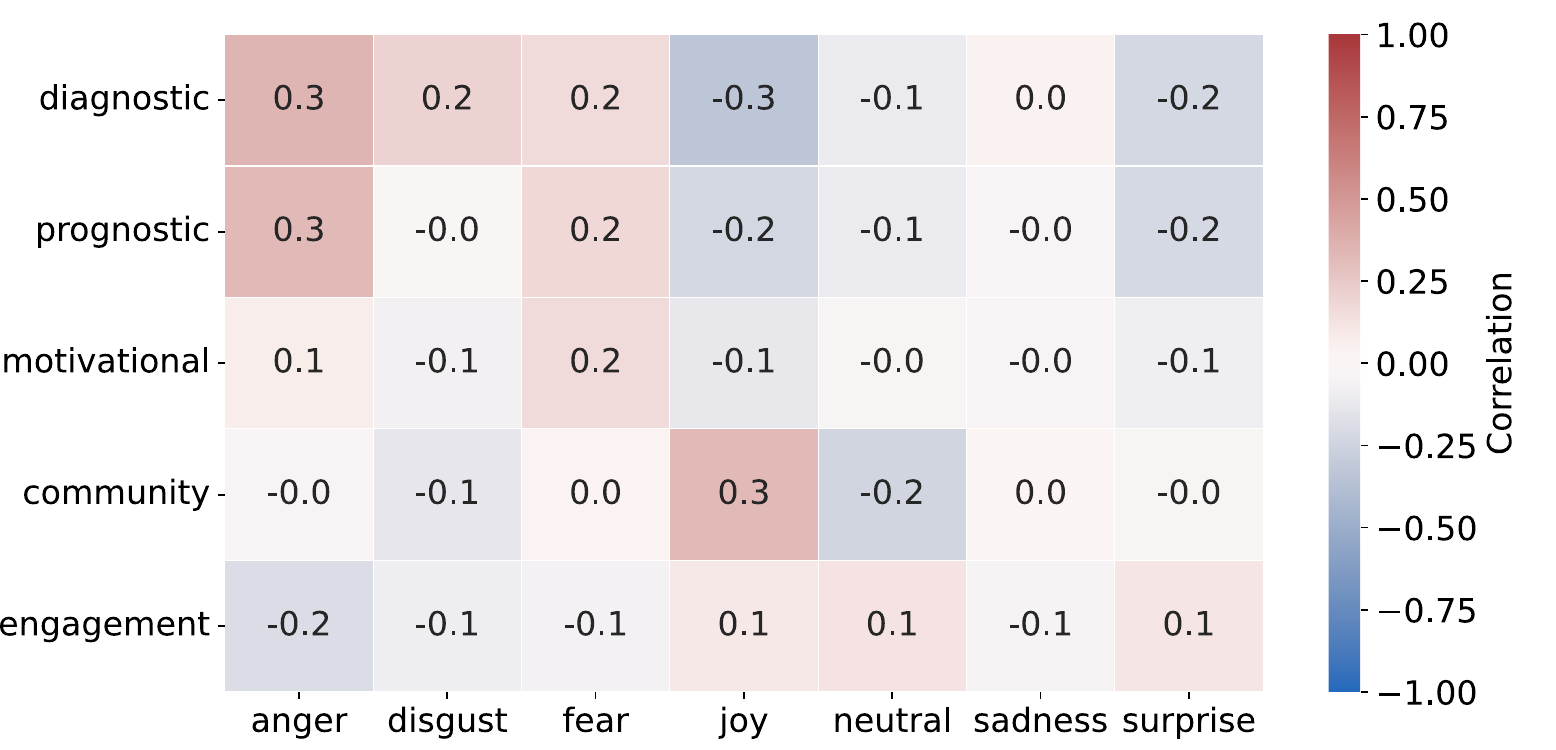}
    \caption{\rev{Pearson's correlation between frames and emotions in Facebook posts. Positive values (red) indicate positive correlations; negative values (blue) indicate negative correlations.}}
    \label{fig:emotion_frames}
\end{figure}

\paragraph{\rev{Emotional Tone}}
\rev{
To explore the affective nuance of frames, we computed correlations between frames and the emotional profiles of posts (cf. Figure~\ref{fig:emotion_frames}).}

\rev{
Diagnostic and prognostic frames were associated with stronger negative emotions, particularly anger and fear, while community frames were moderately correlated with joy. Motivational and engagement frames showed little association with emotions. Overall, correlations were modest, suggesting that emotions provide complementary context rather than replace frames as the primary lens for interpreting union communication strategies.
}

\paragraph{\rev{Topics Discussed}}
\rev{We also explored the thematic content of posts using topic modeling. Applying BERTopic to the full dataset yielded 308 topics with at least 50 posts each. Table~\ref{tab:topic-proportions} reports the ten most prevalent topics (excluding noise).}
\begin{table}[!t]
\centering
\setlength{\tabcolsep}{4pt} 
\begin{tabular}{p{5.7cm} r}
\toprule
\textbf{Topic} & \textbf{\% documents} \\
\midrule
Education and experience of teachers & 3.81 \\
Unions as communities & 2.86 \\
Updates and news from unions & 2.47 \\
Nurses and their importance & 2.04 \\
Congratulations to union members & 1.95 \\
Teamsters events & 1.31 \\
US postal service campaigns & 1.27 \\
Promotion of apprenticeships & 1.03 \\
Promotion of voting during US elections & 0.99 \\
SAG-AFTRA events & 0.96 \\
\bottomrule
\end{tabular}
\caption{\rev{Topic proportions for selected topics, considering all Facebook posts.}}
\label{tab:topic-proportions}
\end{table}

\rev{Some topics, such as \emph{unions as communities} and \emph{promotion of apprenticeships}, are relatively broad and span multiple unions. Others, like \emph{education and experience of teachers} or \emph{nurses and their importance}, are highly union-specific. At the union level, many topics remain highly specific (see Appendix~\ref{app:topic_modeling_unions}). 
These observations indicate that while topic modeling captures meaningful distinctions, prominent themes often reflect the particular context of individual unions. As a result, topic modeling is more useful for qualitative insight than for generalizable inference, and is therefore not included in the predictive model.}

\subsection{Frames Usage Leading to Elections}
\begin{figure}[ht!]
    \centering   
    \includegraphics[width=0.46\textwidth]{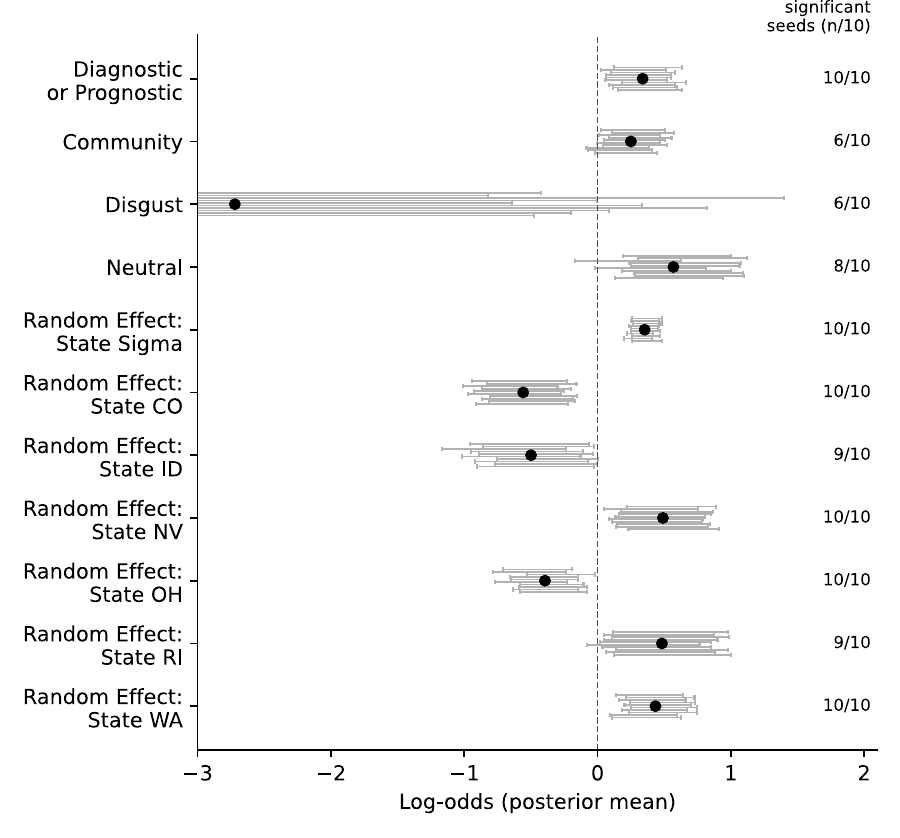}
    \caption{Posterior mean log-odds for selected variables, averaged over ten seeds. Black dots: means; grey lines: 90\% HDIs per seed. Ratios: number of seeds with HDIs excluding zero (shown if $\geq6$ seeds).}
    \label{fig:usage_before}
\end{figure}

\rev{We examined the use of frames in the lead-up to elections and their association with election outcomes using Bayesian logistic regression. To ensure balance, we down-sampled wins and losses within each union and fit the model across ten random seeds. To reduce collinearity, diagnostic and prognostic frame proportions were combined into a single category, as their Variance Inflation Factor (VIF) exceeded 5 when coded separately. Figure~\ref{fig:usage_before} presents posterior mean log-odds estimates for covariates whose posterior distributions excluded zero in at least six out of ten seeds.}

\rev{Results are generally consistent across alternative modeling choices, including separate coding of diagnostic and prognostic frames, a different \rev{moving} aggregation window size, and downsampling of over-represented unions (Figures~\ref{fig:usage_before_onlydiag},~\ref{fig:usage_before_onlyprog},~\ref{fig:usage_before_window3}, and~\ref{fig:usage_before_under90thpercentile}).}

\rev{Greater use of diagnostic/prognostic and community frames prior to elections is associated with higher odds of winning, whereas motivational and engagement frames show no consistent association with the outcome. This pattern suggests that successful campaigns tend to emphasize problem definition, solution-oriented messaging, and collective identity rather than mobilization or engagement cues.}

\rev{With respect to emotional content, neutral language is positively associated with winning, while disgust-related content is negatively associated; however, these effects are less stable across robustness checks than framing effects.}

\rev{State-level random intercepts indicate heterogeneity in baseline odds of winning. Posterior means for Colorado, Idaho, and Ohio are shifted toward lower baseline odds relative to the global intercept, while Nevada, Rhode Island, and Washington exhibit higher baseline odds. The heterogeneity in state-level baseline odds may reflect differences in union density and organizing climate, with states such as Washington and Rhode Island exhibiting stronger labor infrastructures, while lower-density states may pose higher structural barriers to union success.}

\subsection{Frames Usage Pre- and Post-Elections}
\begin{figure}[t!]
    \centering   
    \includegraphics[width=0.47\textwidth]{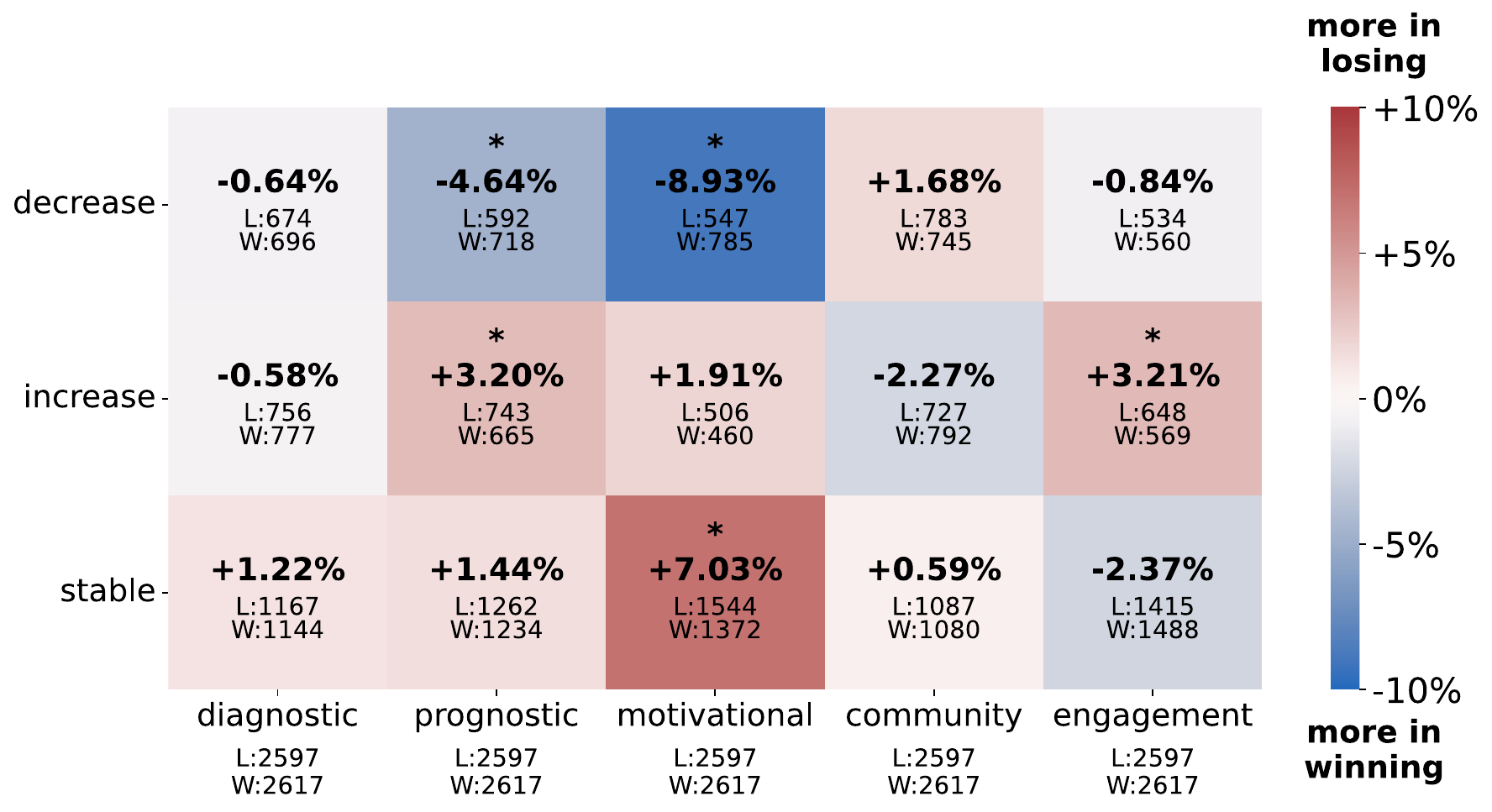}
    \caption{Changes in frame usage pre- vs. post-election, by lost and won cases. Bold values: difference in frame proportions between losses and wins. Counts: union-case instances (L=loss, W=win). $\star$: significant difference ($p<0.05$).}
    \label{fig:usage_before-after}
\end{figure}

We finally examined how frame usage changes from the pre- to the post-election period. 
We first verified that frame usage before and after elections was robust to the choice of \rev{aggregation} window size (see Appendix \ref{app:window_size_rolling}).
For each frame, we classified changes from pre- to post-election into three patterns: decrease, stable, or increase, depending on whether post-election usage was lower, similar, or higher than pre-election usage.
We then counted the number of union–case instances with won or lost outcomes assigned to each pattern, providing an overall measure of how prevalent each change pattern was for each frame.
We down-sampled the number of wins and losses per union to ensure a balanced sample. Results are robust to the choice of sampling seed, as shown in Figure~\ref{fig:usage_before-after_multiseed}.
Figure~\ref{fig:usage_before-after} shows the difference in pattern prevalence between losses and wins, computed as \emph{losing} minus \emph{winning}. Positive values indicate a pattern was more common among losses, negative values indicate it was more common among wins. For example, a value of -5\% in the decrease pattern for the \emph{prognostic} frame means that in won cases, frame usage declined 5\% more often than in lost cases.
We computed Newcombe’s confidence intervals at $\alpha=0.05$ for the difference between the two independent proportions and considered a difference significant if the interval excluded 0.

In won cases, frame usage more often decreased for \emph{prognostic} and \emph{motivational} frames after elections. In lost cases, by contrast, \emph{prognostic} and \emph{engagement} usage increased more often, while \emph{motivational} usage remained stable.

Robustness checks across multiple random seeds and down-sampling of over-represented unions confirmed that the decreases in \emph{prognostic} and \emph{motivational} frames in won cases were consistent. Other patterns were not consistently significant (see Appendix \ref{app:frame_usage_change}).

\section{Discussion and Conclusion}
Traditional sociological research at the intersection of framing and labor movement has often taken a macroscopic perspective when investigating how movement communication relates to organizational outcomes. Such approaches typically conceptualize the labor movement at a collective level. Yet, in the age of connective action~\cite{bennett2012logic}, political action is increasingly dispersed and decentralized~\cite{kavada2016social}, calling for more microscopic perspectives. 
Within social media contexts, this relates to analyzing how individual messages employ discourse frames and connecting these patterns to concrete organizational outcomes, such as labor unions' representation elections.

Some qualitative and mixed-methods studies have already demonstrated the value of this kind of granular analysis~\cite{pasquier2018power, frangi2020tweeting, lazar_mobile_2020}. However, these works tend to be limited to single campaigns, making generalization difficult. Quantitative analyses of union communication, in turn, often examine only a restricted number of accounts, or remain disconnected from organizational outcomes~\cite{hodder2020unions, houghton2021understanding, uba2021political, carneiro2022digital}. Our study addressed these gaps by contributing a micro-level, comparative analysis of 158k Facebook posts by 40 U.S. labor unions between 2015 and 2024. Building on framing theory applied in labor and social movement research~\cite{kelly1998rethinking, benford2000framing, lovejoy2012information, carneiro2022digital}, we developed an annotation codebook and a machine learning model to infer the usage of discourse frames, linking them to representation election outcomes. This approach enabled the kind of systematic, comparative work often missing in framing and social movements scholarship~\cite{tarrow1996social, benford1997insider, snow2014emergence}.

Prior work has documented the untapped mobilization potential of social media for unions and non-profit organizations. Their communication focus is mainly on routine activities, information campaigns, and community-building messages, with very little attention to action-focused messages~\cite{lovejoy2012information, uba2021political, carneiro2022digital}. These studies also highlight the diversity of communication strategies employed across organizations.
Our findings confirmed and extended these insights.

First, we found that nearly 89\% of posts contain at least one frame, with diagnostic and community frames dominating overall discourse (\textbf{RQ1}). This suggests that unions rely heavily on problem identification and community-building to structure their communication. Such practices resonate with the expression of stances on problematic issues, which relate to the diagnostic frame, and celebrations of achievements, common in the community frame, stressed in~\citet{carneiro2022digital}. Moreover, they are close to the relationship-fostering role of social media messaging identified by~\citet{lovejoy2012information}, captured by community frames in our work. In line with such previous work, we found that the use of action-infused social media messages (i.e., our \emph{motivational} frame) was generally very low.
At the same time, substantial heterogeneity emerged across unions, again in line with previous research on diverse communication repertoires. 
Importantly, we observed systematic differences by organizational structure: industrial unions—organizing across skill boundaries—tended to use diagnostic and prognostic frames more than the baseline, whereas craft unions—organized around specialized skill sets—used them less. This pattern resonates with their distinct organizational orientations, with industrial unions more often engaged in large-scale mobilization and bargaining.

Second, \rev{we found that a higher usage of diagnostic, prognostic, and community frames in Facebook communication prior to elections was associated with higher odds of winning} (\textbf{RQ2}).
This finding matches the literature on innovative strategies and union revitalization~\cite{turner2005transformation}, suggesting that successful unions may engage in versatile and intensive framing strategies. It also aligns with research linking frames of injustice and solidarity to mobilization outcomes~\cite{benford2000framing, pasquier2018power}. 

Third, we found dynamic adjustments in the aftermath of elections: won cases more often showed decreases in prognostic and motivational framing compared to lost cases (\textbf{RQ3}). This pattern suggests a shift in priorities once the immediate goal of winning representation is secured, with communication recalibrated from the mobilization to the consolidation phase. Such recalibration aligns with broader theories on organizational life cycles of social movements, considering transitions from mobilization to institutionalization~\cite{christiansen2009four, tarrow2022power}.

\rev{Together}, these findings connect directly to longstanding debates about revitalization of labor unions. Scholars have argued that unions must adapt not only organizationally but also communicatively~\citep{milkman2013back}, experimenting with innovative repertoires~\citep{juravich1998preparing, voss2000breaking}. By systematically distinguishing between the communication practices around successful and unsuccessful election outcomes, our study suggests \rev{that} strategic framing \rev{may} be an important part of these adaptation processes. More broadly, we show the value of micro-level, comparative framing analysis for advancing our understanding of how movements strategize and evolve in the digital era.
\rev{It is important to note, however, that these conclusions are drawn from observational, correlational analyses and predictive modeling; as such, they do not support causal claims about the effects of specific framing strategies on election outcomes.}

\paragraph{Limitations, Implications and Future Work}
While our study makes both methodological and substantive contributions, several limitations should be noted. 
\rev{First, our data is observational and correlational; consequently, observed associations between communication features and electoral success cannot be interpreted causally.}
\revii{In particular, framing may reflect underlying organizational capacity rather than independent strategic choice: unions with greater resources, stronger leadership, or more intensive offline organizing may both communicate differently and be more likely to win elections.}
Second, the analysis is limited to Facebook, an important but partial element of unions’ communication strategies, which also involve other digital platforms and offline channels. \rev{Out of over 500 unions with available election outcomes, only 66 had Facebook accounts, and only 40 provided downloadable data, resulting in a relatively small and potentially skewed sample. Consequently, our findings may not generalize to all unions, particularly those for which Facebook is not a primary communication channel}. 
\rev{Third, our analysis focuses on posts from the week immediately before and after each election.}
\revii{While this window is intended to capture communication tied to electoral events, it may be too narrow to reflect longer-term framing strategies or too broad, capturing stable communication patterns rather than election-specific adjustments. To mitigate the latter concern, we account for longer-term trends in frame usage by leveraging periods without elections.}
\rev{Fourth}, we focus on national and international level union accounts to ensure sufficient activity, which may obscure dynamics in smaller or more localized organizations. \rev{Changes in data access policies could allow for analyses that can estimate the impact of such issues.}
\rev{Fifth}, we analyzed five discourse frames identified in prior work on social movements and civil society organizations. This enhances comparability but may overlook additional frames or nuances. 
Finally, unions differ in size, resources, and political context, factors likely shaping communication in ways not fully accounted for here.
We also acknowledge that systematic analyses of union communication could be misused by \rev{political actors, employers, or other organizations, underscoring the need for careful ethical reflection when applying these methods.}

Despite these limitations, the study has implications for multiple audiences. For labor unions, the findings can inform digital communication strategies, highlighting how different frames can be used in election campaigns. For political communication researchers, the study offers large-scale empirical evidence on how civil society organizations employ framing online. For organizations more broadly, our results suggest ways to refine messaging to balance mobilization, solidarity, and engagement objectives. Methodologically, the released classifier and dataset provide a scalable resource for studying digital discourse framing across movements.

Future work could extend this study in several directions. Expanding the frame taxonomy would allow more fine-grained analyses of communication strategies, including issue-specific frames.  
Examining union accounts at smaller, local levels—beyond the national and international accounts analyzed here—could reveal additional variation in messaging.
Multi-platform analyses would provide a more holistic picture of unions’ communication ecosystems beyond Facebook. Incorporating a perspective on the targets of unions' communication would also provide a broader understanding of strategic efforts and complement existing literature \rev{that suggests} ad hoc targeting as one valuable revitalization strategy~\cite{earl2019symposium, uba2021political}.  
Finally, incorporating audience responses through engagement metrics and comments would help link message production with resonance, offering deeper insight into the interplay between framing practices and collective action dynamics.

\section{Related Work}
\subsection{NLRB Election Data}
The National Labor Relations Board (NLRB) election data has been widely used to study unionization dynamics in the United States, particularly predictors of union success.
Early work by~\citet{maranto1987effect} highlights union size, internal democracy, strike activity, and centralized decision-making as key determinants of certification outcomes. Extending this perspective,~\citet{ferguson2008eyes} models organizing as a sequential process from petitions to first contracts, highlighting union size and unfair labor practice charges as critical across stages.

A second strand of studies combines NLRB data with external sources to investigate how broader institutional and demographic contexts shape unionization. For example,~\citet{sojourner2022effects} connects certification to workplace safety enforcement, while~\citet{ferguson2015control} studies workplace composition using Equal Employment Opportunity Commission data.~\citet{bronfenbrenner2005organizing} takes a related approach by comparing bargaining units across gender compositions, assessing the role of company characteristics, union demographics, and organizing tactics.

Other studies situate elections within broader political-economic dynamics.~\citet{stepan2010rival} analyzes inter-union competition, while~\citet{schaller2023decomposing} links union decline to sectoral and regional shifts. Finally,~\citet{ferguson2018osmotic} connects labor organizing to wider social movements, introducing the concept of \emph{osmotic mobilization} to demonstrate how protest activity in society at large correlates with support for unions in representation elections.

This literature demonstrates the versatility of NLRB election data for analyzing both the internal dynamics of organizing and its intersection with external institutional, demographic, and political factors. 
While prior work focuses on predictors of success, workplace outcomes, and long-term trends, we incorporate the digital domain, using NLRB data to anchor the analysis of online communication.

\subsection{Digital Communication of Labor and Social Movements}

\subsubsection{Labor Movement}
With the widespread adoption of digital technologies, labor unions increasingly use online platforms to engage members, recruit participants, and communicate with stakeholders. Union strategies are shaped by both external political contexts and internal organizational choices~\cite{martinez2003new}. Among revitalization strategies, \emph{organizing}—recruiting, representing, retaining members, and expanding mobilization capacity—is particularly prominent in digital settings~\cite{frege_2004_varieties}. Communication plays a central role in these efforts, as discourse frames help articulate grievances, propose solutions, and mobilize collective identities.

A substantial body of work links framing to labor organizing. 
Building on collective action theory~\cite{snow1988ideology},~\citet{kelly1998rethinking} argues that successful mobilization depends on workers perceiving injustice, identifying solutions, and believing in collective efficacy.
Similarly,~\citet{voss2000breaking} highlights the importance of identifying the right strategic repertoires of communication and engagement for the labor revitalization efforts to be successful. Other foundational work that followed highlights the importance of identifying the right frames for women and minorities~\cite{bronfenbrenner2005organizing}, and millenials~\cite{milkman2013back}.

Research on digital communication has largely focused on mobilization. Studies of the Fight for Fifteen movement show how social media supports mobilization and connects collective action with grassroots dynamics~\cite{pasquier2018power, pasquier2020towards, frangi2020tweeting}. Other work analyzes mobilizing frames across platforms~\cite{houghton2021understanding}, coding social media posts in terms of framing, attribution, and calls to action. Beyond mobilization, research has also examined how digital networks support organizational purposes, such as building collective voices prior to mobilization~\cite{lazar_mobile_2020}, or reaching marginalized groups otherwise excluded from traditional organizing channels~\cite{wood2015networks, thornthwaite2018unions}. 
The effectiveness of communication strategies is also contingent on other important factors, such as organizational and leadership structures, as demonstrated in the comparative analysis of 2018 teachers' strikes in Oklahoma and Arizona~\citep{blanc2022digitized}.

Despite these insights, evidence suggests that unions are not fully exploiting the potential of digital media. ~\citet{hodder2020unions} finds that online campaigning and news reporting remain dominant uses, leaving recruitment opportunities largely untapped. Likewise,~\citet{carneiro2022digital} documents outdated strategies centered on one-way information sharing, rather than fostering engagement. More promising results emerge when unions tailor their communication to specific worker groups~\cite{jansson2019trade} and account for the diversity of audiences with distinct communication needs~\cite{panagiotopoulos_digital_2021}. These targeted strategies appear to be a necessary step towards building more effective digital communication. 

\subsubsection{Beyond the Labor Movement}
Beyond unions, a large body of research examines how social movements and civil society organizations use digital platforms to structure collective action and mobilize supporters. 

One line of work emphasizes the role of \emph{mobilization frames} in online environments. Building on framing theory, scholars show how social media enables the diffusion of diagnostic, prognostic, and motivational frames, crucial for lowering barriers to participation~\cite{benford2000framing, fernandez2023digital, mendelsohn2024framing, pera2025extracting}. 
A second perspective highlights the shift from the \emph{logic of collective action} toward the \emph{logic of connective action}~\cite{bennett2012logic}. While traditional mobilization relies on formal organizations and collective identities, connective action underscores personalized, digitally mediated participation that can occur without strong organizational structures. This framework has been applied to cases such as the Arab Spring~\cite{steinert2017spontaneous} and climate activism~\cite{segerberg2011social}. Finally, research on digital \emph{affordances} investigates how platform features—such as sharing content, meta-voicing, and multimedia storytelling—shape the opportunities and constraints of collective action~\cite{zheng2016affordances, saebo2020combining}.

\vspace{0.3cm}
\noindent While prior research has provided valuable insights into social media mobilization and organizational communication, many studies rely on qualitative analyses or selected use cases. In contrast, our study leverages a large-scale computational approach to systematically examine union social media communication during times of representation elections at the micro-level of individual messages. Moreover, by linking communication patterns to NLRB election outcomes, we explore how online framing relates to tangible organizational success, providing empirical evidence on the functional value of digital strategies that complements existing theoretical and qualitative work.

\section*{Acknowledgements}
This work was supported by the Carlsberg Foundation through the COCOONS project (CF21-0432), and partially supported by the National Science Foundation (Grant IIS-1815875). Funders had no role in study design, data collection and analysis, decision to publish, or preparation of the manuscript.

\bibliography{aaai25}

\clearpage

\section{Paper Checklist}

\begin{enumerate}

\item For most authors...
\begin{enumerate}
    \item  Would answering this research question advance science without violating social contracts, such as violating privacy norms, perpetuating unfair profiling, exacerbating the socio-economic divide, or implying disrespect to societies or cultures?
    \answerYes{Yes.}
  \item Do your main claims in the abstract and introduction accurately reflect the paper's contributions and scope?
    \answerYes{Yes.}
   \item Do you clarify how the proposed methodological approach is appropriate for the claims made? 
    \answerYes{Yes, in the ``Methodological Framework" section.}
   \item Do you clarify what are possible artifacts in the data used, given population-specific distributions?
    \answerYes{Yes, we discuss potential biases in our data in the ``Discussion and Conclusion'' section.}
  \item Did you describe the limitations of your work?
    \answerYes{Yes, limitations are presented and discussed in the ``Discussion and Conclusion'' section.}
  \item Did you discuss any potential negative societal impacts of your work?
    \answerYes{Yes, we address negative societal impact in the ``Discussion and Conclusion'' section.}
      \item Did you discuss any potential misuse of your work?
    \answerYes{Yes, we discuss potential misuse in the ```Discussion and Conclusion'' section.}
    \item Did you describe steps taken to prevent or mitigate potential negative outcomes of the research, such as data and model documentation, data anonymization, responsible release, access control, and the reproducibility of findings?
    \answerYes{Yes, we share the steps and our framework in the ``Methodological Framework" section. Full experimental details (code) are provided as a shared repository.}
  \item Have you read the ethics review guidelines and ensured that your paper conforms to them?
    \answerYes{Yes.}
\end{enumerate}

\item Additionally, if your study involves hypotheses testing...
\begin{enumerate}
  \item Did you clearly state the assumptions underlying all theoretical results?
    \answerYes{Yes, we frame our experiments within the scope of previous research in the ``Theoretical Framework'', ``Methodological Framework", and ``Related Work" sections.}
  \item Have you provided justifications for all theoretical results?
    \answerYes{Yes, justifications are provided in the ``Results" and ``Discussion and Conclusion" sections.}
  \item Did you discuss competing hypotheses or theories that might challenge or complement your theoretical results?
    \answerYes{Yes, other approaches in terms of theoretical frameworks and the motivation behind our choice are described in the ``Methodological Framework" and the ``Related Work" sections.}
  \item Have you considered alternative mechanisms or explanations that might account for the same outcomes observed in your study?
    \answerYes{Yes, in the ``Results" and ``Discussion and Conclusion" sections we explore potential confounders of our results and highlight limitations of our work.}
  \item Did you address potential biases or limitations in your theoretical framework?
    \answerYes{Yes, these are addressed in the ``Discussion and Conclusion" section.}
  \item Have you related your theoretical results to the existing literature in social science?
    \answerYes{Yes, we directly relate our work to previous work in ``Introduction", ``Related Work", and ``Theoretical Framework''.}
  \item Did you discuss the implications of your theoretical results for policy, practice, or further research in the social science domain?
    \answerYes{Yes, these are discussed in the ``Discussion and Conclusion" section.}
\end{enumerate}

\item Additionally, if you are including theoretical proofs...
\begin{enumerate}
  \item Did you state the full set of assumptions of all theoretical results?
    \answerNA{NA.}
	\item Did you include complete proofs of all theoretical results?
    \answerNA{NA.}
\end{enumerate}

\item Additionally, if you ran machine learning experiments...
\begin{enumerate}
  \item Did you include the code, data, and instructions needed to reproduce the main experimental results (either in the supplemental material or as a URL)?
    \answerYes{Yes. Included link to shared repository (see ``Introduction''). Configuration details are reported in Appendix.}
  \item Did you specify all the training details (e.g., data splits, hyperparameters, how they were chosen)?
    \answerYes{Yes, whenever necessary. Described in ``Methodological Framework" and Appendix. Details are given in the code provided in the shared repository.}
     \item Did you report error bars (e.g., with respect to the random seed after running experiments multiple times)?
    \answerYes{Yes, whenever relevant. Specifically, we took into account confidence intervals is Figures \ref{fig:usage_before} and \ref{fig:usage_before-after} in the main text and in robustness checks reported in the Appendix.}
    \item Did you include the total amount of compute and the type of resources used (e.g., type of GPUs, internal cluster, or cloud provider)?
    \answerYes{Yes, see reference to computational resources in ``Methodological Framework'' and details in the Appendix.}
     \item Do you justify how the proposed evaluation is sufficient and appropriate to the claims made? 
    \answerYes{Yes, specified in the ``Methodological Framework" and ``Results" section.}
     \item Do you discuss what is ``the cost`` of misclassification and fault (in)tolerance?
    \answerYes{Yes, in the ``Results" section.}
  
\end{enumerate}

\item Additionally, if you are using existing assets (e.g., code, data, models) or curating/releasing new assets, \textbf{without compromising anonymity}...
\begin{enumerate}
  \item If your work uses existing assets, did you cite the creators?
    \answerYes{Yes, creators of models used as part of the experiments are cited in ``Methodological Framework" section.}
  \item Did you mention the license of the assets?
    \answerYes{Yes, see Appendix.}
  \item Did you include any new assets in the supplemental material or as a URL?
    \answerYes{Yes, data and code are included in the shared repository linked in the ``Introduction''.}
  \item Did you discuss whether and how consent was obtained from people whose data you're using/curating?
    \answerNA{NA.}
  \item Did you discuss whether the data you are using/curating contains personally identifiable information or offensive content?
    \answerYes{Our data contains PII of labor unions in the United States. We discuss this in the DataSheet (Supplementary Material).}
\item If you are curating or releasing new datasets, did you discuss how you intend to make your datasets FAIR?
\answerYes{Yes, see DataSheet in the Supplementary Material.}
\item If you are curating or releasing new datasets, did you create a Datasheet for the Dataset (see~\citet{gebru2021datasheets})? 
\answerYes{Yes, see the Supplementary Material.}
\end{enumerate}

\item Additionally, if you used crowdsourcing or conducted research with human subjects, \textbf{without compromising anonymity}...
\begin{enumerate}
  \item Did you include the full text of instructions given to participants and screenshots?
     \answerNA{NA.}
  \item Did you describe any potential participant risks, with mentions of Institutional Review Board (IRB) approvals?
     \answerNA{NA.}
  \item Did you include the estimated hourly wage paid to participants and the total amount spent on participant compensation?
     \answerNA{NA.}
   \item Did you discuss how data is stored, shared, and deidentified?
    \answerNA{NA.}
\end{enumerate}

\end{enumerate}

\clearpage

\appendix

\setcounter{figure}{0}
\setcounter{table}{0}
\setcounter{equation}{0}
\renewcommand{\thefigure}{C\arabic{figure}}
\renewcommand{\thetable}{C\arabic{table}}
\renewcommand{\theequation}{C\arabic{equation}}

\setcounter{secnumdepth}{2}

\section{Frames Annotation} \label{app:annotation_codebook}
We annotated Facebook posts for the presence of five discourse frames: \emph{diagnostic}, \emph{prognostic}, \emph{motivational}, \emph{community}, and \emph{engagement}. In addition, given the close ties between labor unions and political events (e.g., presidential elections, support for congressional candidates), we also annotated posts with an auxiliary label: \emph{political endorsement}. This label captures instances in which unions explicitly support political actors or candidates.  

The complete annotation guidelines are provided in the supplementary material (\texttt{codebook.pdf}).

\section{Modeling and Computational Details} \label{app:modeling_resources}
We fine-tuned a RoBERTa-base model to capture the characteristics of unions' Facebook communications. For this fine-tuning of the language model, we used 5 epochs with a batch size of 128, keeping all other parameters at their default values. Training was performed on a Tesla V100 GPU and required about 30 minutes. 

For the multi-label classification task, we trained the RoBERTa model for 30 epochs with a learning rate of $4 \times 10^{-5}$ and a batch size of 32. To address class imbalance, we applied class weights computed as  
\[
\frac{\text{total samples}}{2 \times [n_{\text{class 0}}, \ldots, n_{\text{class n}}]}.
\]  
All other parameters were left at default settings. Training required a few minutes on a Tesla V100.  

\rev{We also evaluated a BERT-base uncased model that underwent the same fine-tuning pipeline and resulted in similar computational costs.}

The multi-label classification setup included the five main discourse frames and the additional \emph{political endorsement} label. However, given the very limited presence of political endorsement in the annotated dataset (3\% of posts across training and test sets) and in the full Facebook dataset (2.5\% of posts), we did not retain this label in the main analysis. We therefore focus exclusively on the five core discourse frames in the paper.

\subsection{Licenses}
The \texttt{RoBERTa base} model used for building the multi-label communication frames classifier is licensed under a MIT license. \rev{The \texttt{BERT-base} model used as a comparison is licensed under an Apache License, Version 2.0.
The emotion classification model is licensed under an Apache License, Version 2.0.}

\section{\rev{Topic Modeling on Selected Unions}} \label{app:topic_modeling_unions}
\begin{figure}[!ht]
    \centering

    \begin{subfigure}{0.47\textwidth}
        \includegraphics[width=\linewidth]{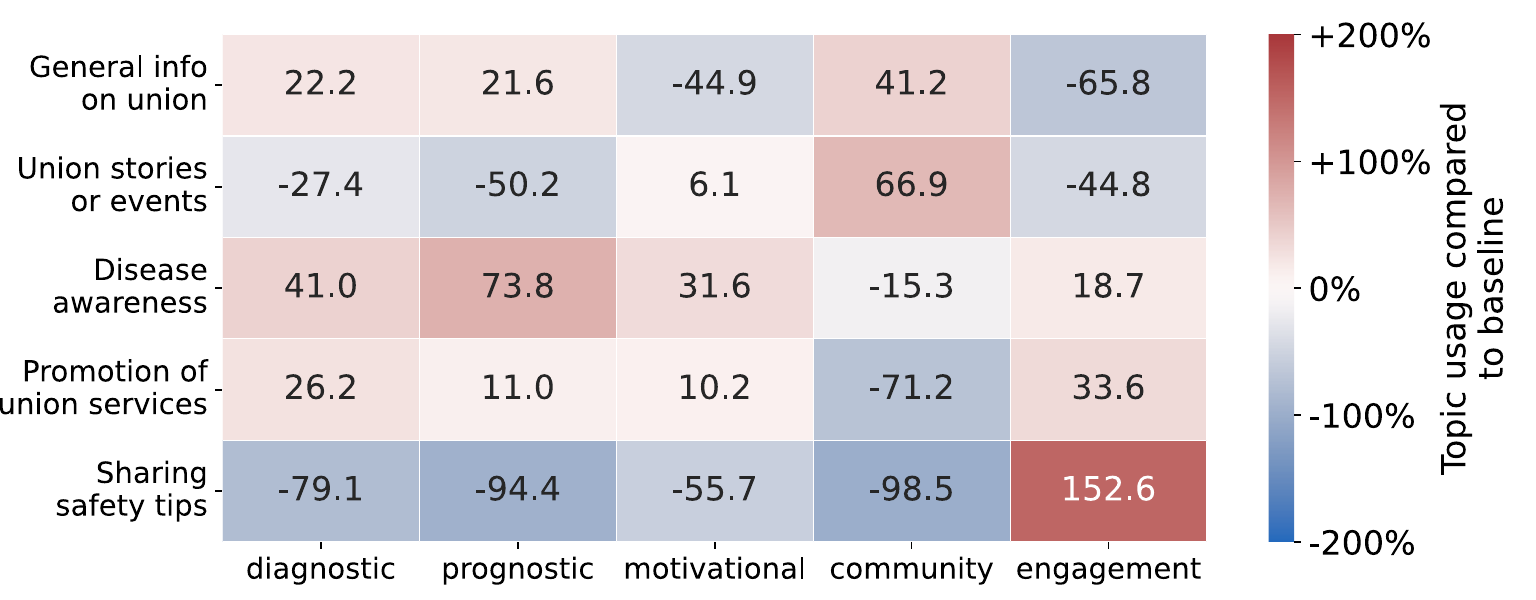}
        \caption{IAFF (Craft)}
    \end{subfigure}

    \par\bigskip 

    \begin{subfigure}{0.47\textwidth}
        \includegraphics[width=\linewidth]{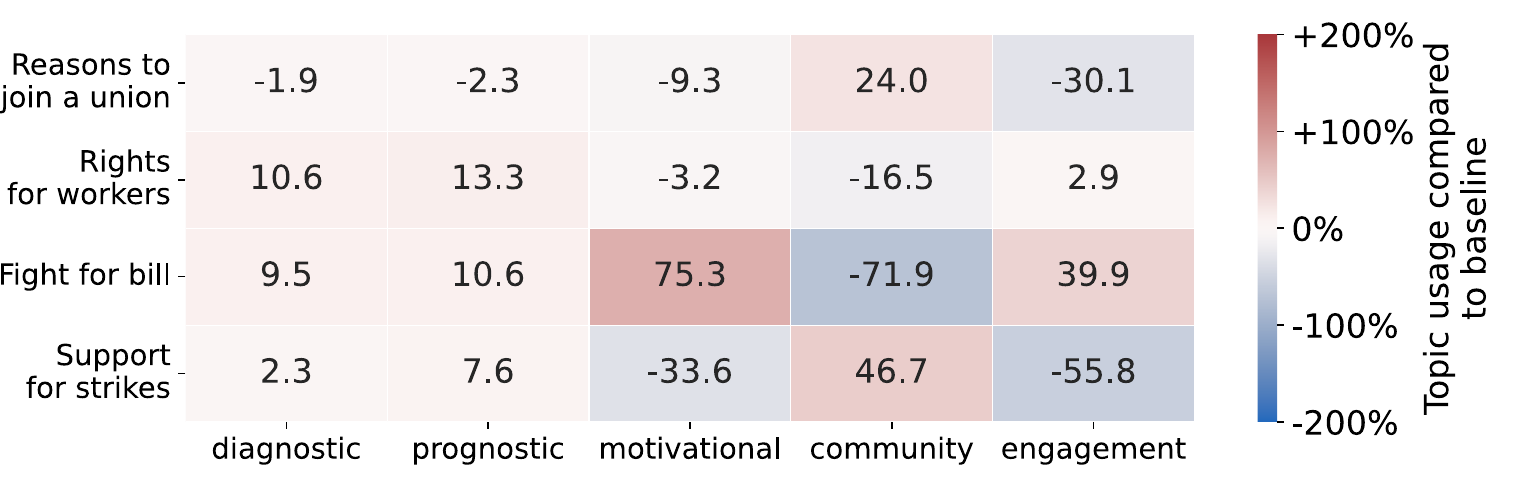}
        \caption{SEIU (Industrial)}
    \end{subfigure}

    \caption{\rev{Usage of topics within posts with a given frame relative to overall topic usage. Cell values represent percentages; red indicates higher-than-baseline usage, blue indicates lower-than-baseline usage.}}
    \label{fig:topics_frames}
\end{figure}

\rev{To explore the relationship between topics and frames, we examined posts from two of the most active (i.e., by total number of posts) craft and industrial unions on Facebook, namely IAFF and SEIU.
Using topic modeling through BERTopic, we analyzed their posts in terms of prevalent topics. Figure~\ref{fig:topics_frames} presents manually merged, overarching topics that appear in at least 5\% of posts, showing how the prevalence of each topic within a frame deviates from its overall usage.}

\rev{In the main text, we pointed at the fact that applying topic modeling across all posts resulted in many union-specific topics.
Even at the union level, some topics remain highly specific---for example, \emph{sharing safety tips} for IAFF. Certain topics align with particular frames: in IAFF, \emph{union stories or events} is more common in posts using the \emph{community} frame, while \emph{sharing safety tips} appears more often in the \emph{engagement} frame. In SEIU, \emph{reasons to join a union} and \emph{support for strikes} are more often associated with \emph{community} frames. At the same time, many topics, such as IAFF’s \emph{general info on union} or SEIU’s \emph{rights for workers} and \emph{fight for bill}, cut across frames without a single dominant association.}

\rev{Overall, topic modeling is useful for capturing additional nuance within frames, highlighting the specific issues discussed. However, its highly union-specific nature limits its usefulness for generalizable predictions beyond frame-level analysis.}

\section{Robustness Checks} \label{app:robustness_checks}
\subsection{\rev{Alternative Classifiers}}
\label{app:alt_classifier}
\setcounter{figure}{0}
\setcounter{table}{0}
\setcounter{equation}{0}
\renewcommand{\thefigure}{D\arabic{figure}}
\renewcommand{\thetable}{D\arabic{table}}
\renewcommand{\theequation}{D\arabic{equation}}

\rev{We considered two alternative classifiers in to compare with the RoBERTa-base one fine-tuned on language and then on labeled data, referenced in the main text.
The first is a RoBERTa-base classifier simply fine-tuned on the labeled data, skipping the language fine-tuning step.
We report its performance on the test set in Table \ref{tab:classification-report_robertav2}.}
\setlength{\tabcolsep}{5pt} 
\begin{table}[ht]
\centering
\begin{tabular}{@{}lcccc@{}}
\toprule
\textbf{Frame} & \textbf{Precision} & \textbf{Recall} & \textbf{F1-score} & \textbf{Support} \\
\midrule
Diagnostic            & 0.77 & 0.86 & 0.81 & 84 \\
Prognostic            & 0.70 & 0.72 & 0.71 & 58 \\
Motivational          & 0.59 & 0.83 & 0.69 & 29 \\
Community             & 0.67 & 0.72 & 0.69 & 81 \\
Engagement            & 0.72 & 0.82 & 0.77 & 51 \\
\midrule
Micro avg             & 0.70 & 0.79 & 0.74 & 303 \\
Macro avg             & 0.69 & 0.79 & 0.73 & 303 \\
Weighted avg          & 0.70 & 0.79 & 0.74 & 303 \\
\bottomrule
\end{tabular}
\caption{\rev{Performance of the multi-label discourse frames classifier (RoBERTa-base, just fine-tuned for the classification task)}.}
\label{tab:classification-report_robertav2}
\end{table}

\rev{The second is a BERT-base uncased model, fine-tuned on language across all Facebook posts and then on the labeled data for multi-label classification. We report its performance in Table \ref{tab:classification-report_bert}.}
\setlength{\tabcolsep}{5pt} 
\begin{table}[ht]
\centering
\begin{tabular}{@{}lcccc@{}}
\toprule
\textbf{Frame} & \textbf{Precision} & \textbf{Recall} & \textbf{F1-score} & \textbf{Support} \\
\midrule
Diagnostic            & 0.80 & 0.82 & 0.81 & 84 \\
Prognostic            & 0.67 & 0.62 & 0.64 & 58 \\
Motivational          & 0.58 & 0.66 & 0.61 & 29 \\
Community             & 0.62 & 0.65 & 0.63 & 81 \\
Engagement            & 0.62 & 0.67 & 0.64 & 51 \\
\midrule
Micro avg             & 0.67 & 0.70 & 0.68 & 303 \\
Macro avg             & 0.66 & 0.68 & 0.67 & 303 \\
Weighted avg          & 0.67 & 0.70 & 0.68 & 303 \\
\bottomrule
\end{tabular}
\caption{\rev{Performance of the multi-label discourse frames classifier (BERT-base uncased, fine-tuned for language and then fine-tuned for the classification task)}.}
\label{tab:classification-report_bert}
\end{table}

\subsection{\rev{Coding Diagnostic and Prognostic in a Single Category}} \label{app:diagprog}
\begin{figure}[ht!]
    \centering   
    \includegraphics[width=0.43\textwidth]{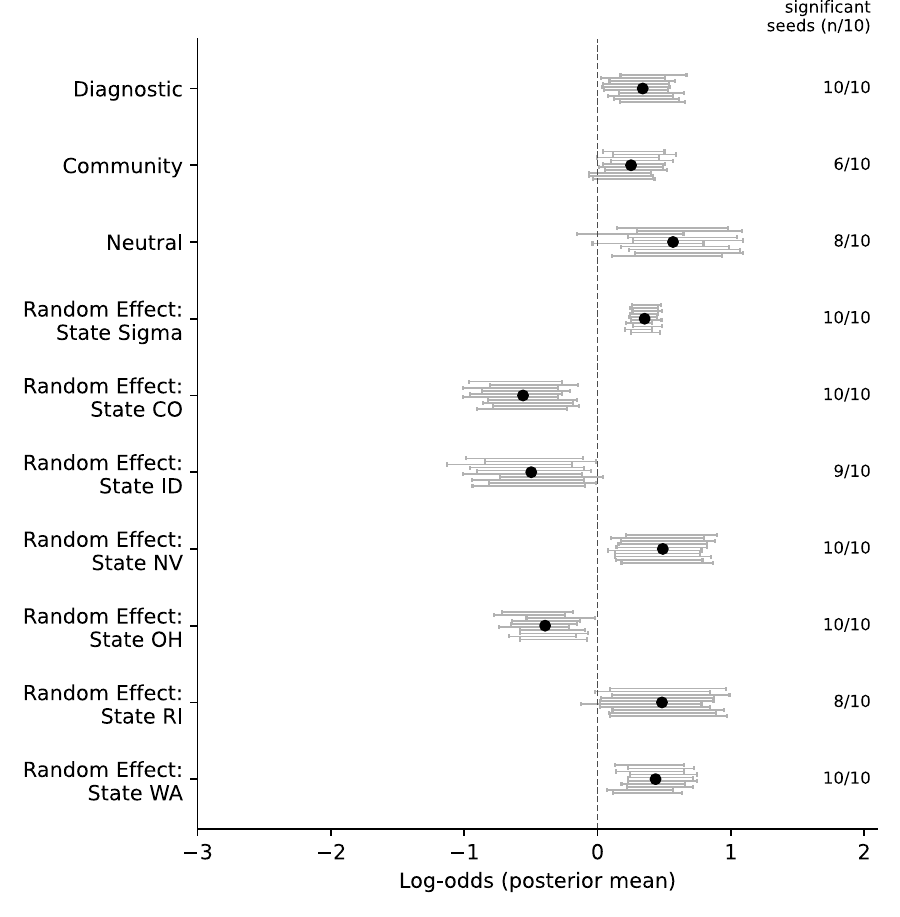}
    \caption{\rev{Posterior mean log-odds estimates across ten seeds, not considering the proportion of prognostic frames in the predictors. Variables not shown had no credible association with the outcome or credible association for fewer than six seeds.}}
    \label{fig:usage_before_onlydiag}
\end{figure}

\begin{figure}[ht!]
    \centering   
    \includegraphics[width=0.43\textwidth]{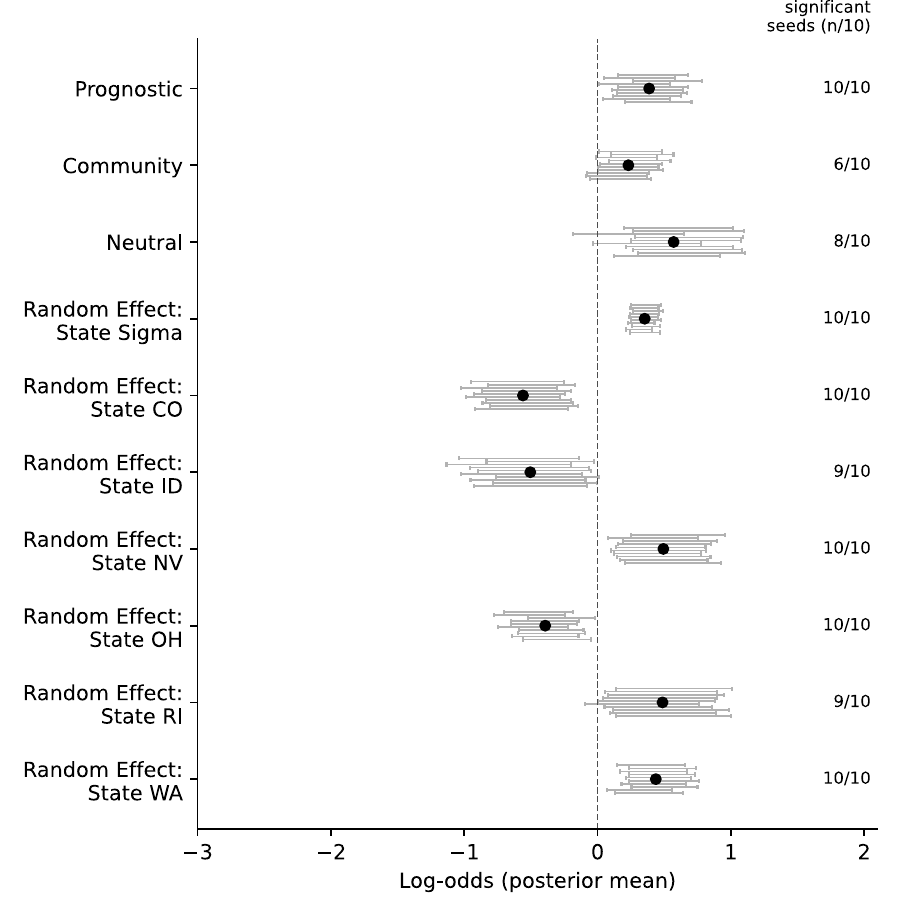}
    \caption{\rev{Posterior mean log-odds estimates across ten seeds, not considering the proportion of diagnostic frames in the predictors. Variables not shown had no credible association with the outcome or credible association for fewer than six seeds.}}
    \label{fig:usage_before_onlyprog}
\end{figure}

\rev{The predictive power of frames over election winning is robust to the coding of diagnostic and prognostic into a single category rather than considering them as separate predictors. The only exception is disgust, whose posterior distribution excludes zero in fewer than six out of ten model runs when diagnostic and prognostic frames are modeled separately, indicating sensitivity to model specification rather than a robust association. Figures \ref{fig:usage_before_onlydiag} and \ref{fig:usage_before_onlyprog} illustrate this pattern for models including only the proportion of diagnostic and prognostic posts prior to the election, respectively.
}

\subsection{Window Size for Aggregation} \label{app:window_size_rolling}
\begin{figure}[ht!]
    \centering   
    \includegraphics[width=0.43\textwidth]{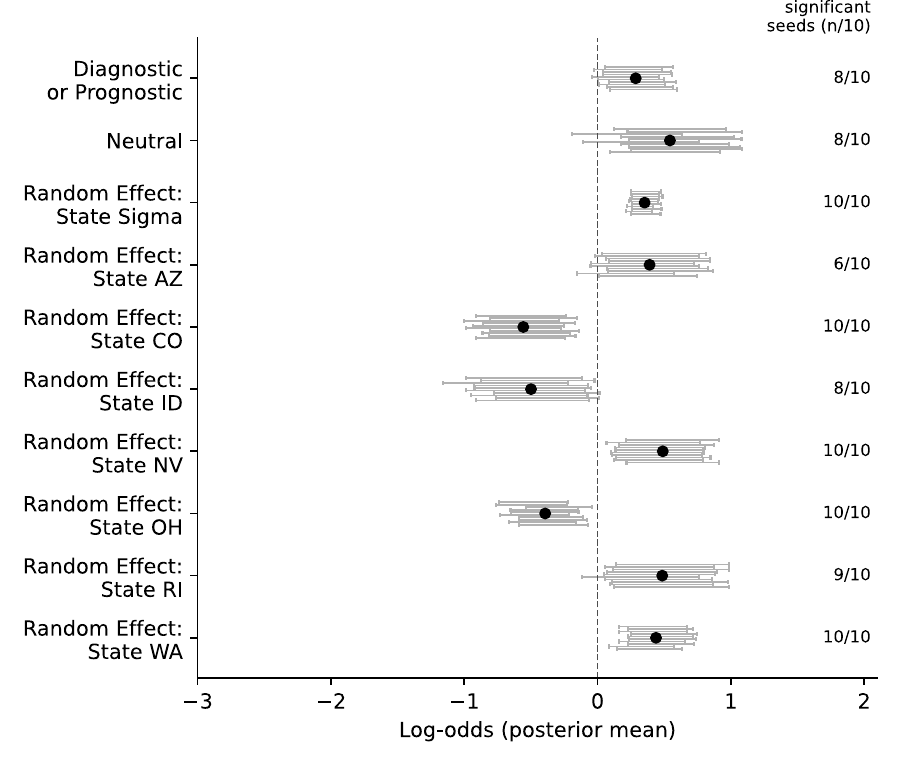}
    \caption{\rev{Posterior mean log-odds estimates across ten seeds, considering a window of three days for aggregation. Variables not shown had no credible association with the outcome or credible association for fewer than six seeds.}}
    \label{fig:usage_before_window3}
\end{figure}

\rev{The estimated effects of diagnostic and prognostic frames on election outcomes remain stable across alternative aggregation window sizes. In contrast, the posterior distribution for community framing increasingly overlaps zero, indicating reduced evidence for an association under this specification. The estimated effect of neutral emotional content remains robust across window sizes. Random intercepts for states are largely stable, with the exception of Arizona, whose posterior distribution is shifted toward higher baseline odds of winning in this configuration (cf. Figure \ref{fig:usage_before_window3}).}

\subsection{Imbalance for Number of Cases per Union} \label{app:imbalance_n_cases}

\subsubsection{Frames Usage Leading to Elections} \label{app:frame_usage_before}

\rev{In the main paper, we already demonstrated the robustness of coefficient estimates to different random seeds used to balance wins and losses within unions.}

\rev{We next assess robustness to variation in the number of representation cases per union. Specifically, we capped the total number of cases per union at the 90th percentile of the distribution (considering winning and losing outcomes jointly) and repeated the analyses across ten random seeds. For each seed, we first balanced the number of wins and losses per union and then fitted a Bayesian logistic regression model predicting the binary election outcome.}

\begin{figure}[!t]
    \centering   
    \includegraphics[width=0.43\textwidth]{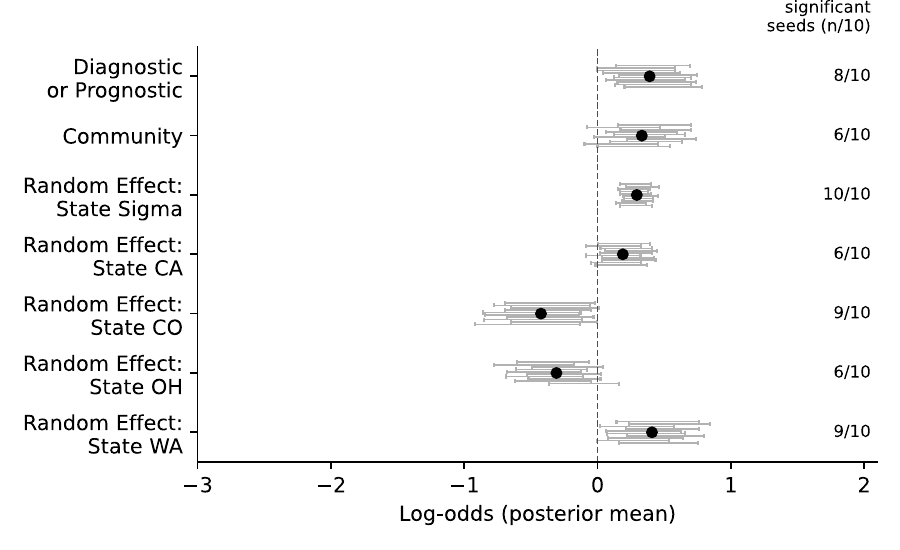}
    \caption{\rev{Posterior mean log-odds estimates across ten seeds, downsampling over-represented unions.}}
    \label{fig:usage_before_under90thpercentile}
\end{figure}

\rev{Figure~\ref{fig:usage_before_under90thpercentile} presents posterior log-odds estimates for selected model variables. Overall, the results are robust to differences in the number of representation cases across unions. Estimated effects of diagnostic/prognostic and community frames remain stable, whereas posterior distributions for disgust-related and emotionally neutral content increasingly overlap zero under this specification. State-level random intercepts are largely consistent across models, with Colorado, Ohio, and Washington exhibiting similar deviations from the global intercept, while California’s posterior distribution is shifted toward higher baseline odds of winning.}

\subsubsection{Frames Usage Pre- and Post-Elections} \label{app:frame_usage_change}
\begin{figure}[!t]
    \centering   
    \includegraphics[width=0.45\textwidth]{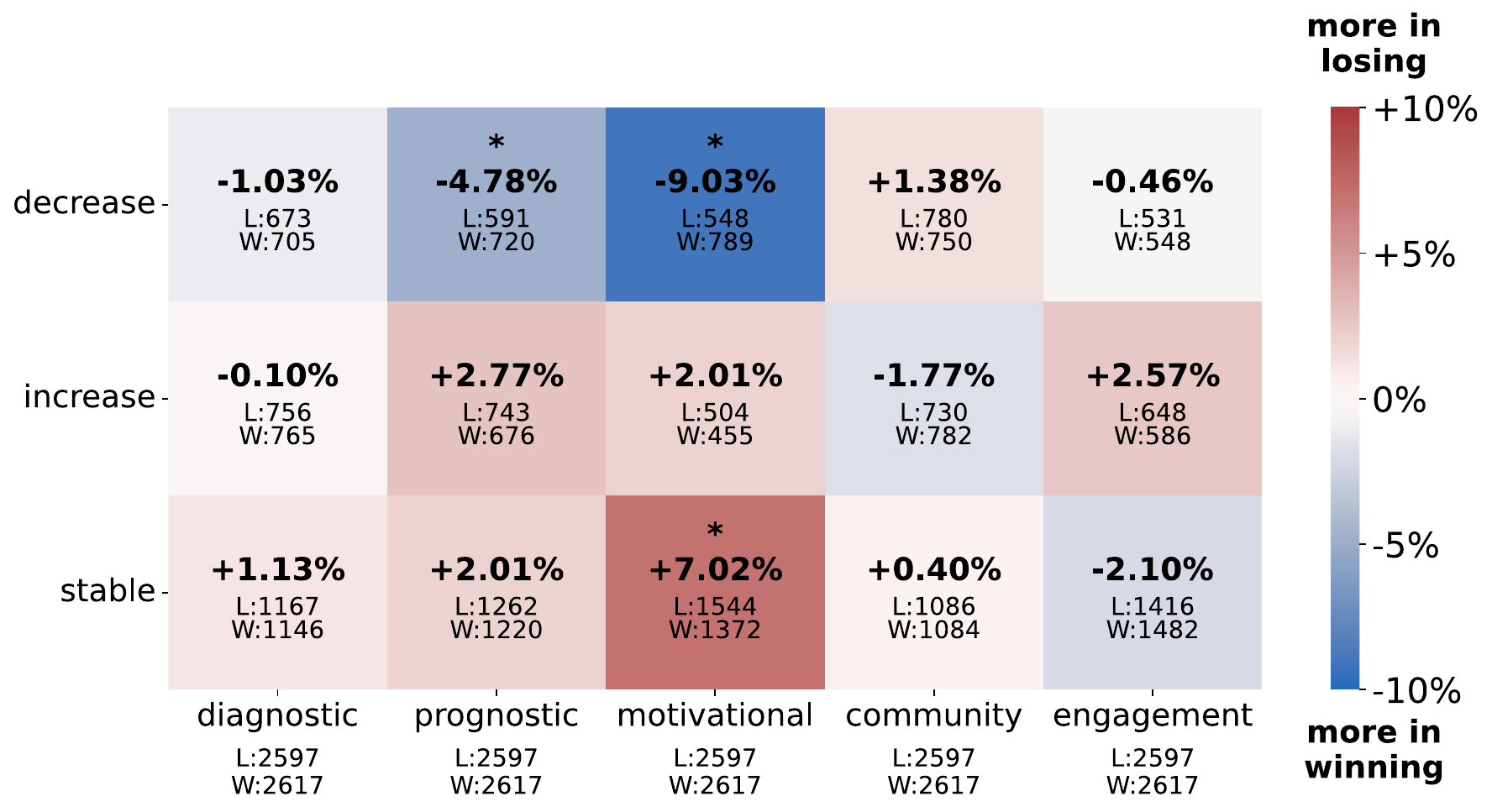}
    \caption{Usage of frames before and after the event, comparing lost and won cases. 
    The $\star$ sign indicates a significant difference between outcome types \rev{($p < 0.05$)}.}
    \label{fig:usage_before-after_multiseed}
\end{figure}

Given that we balanced the number of wins and losses per union, we verified the robustness of results to the choice of random seed for sampling. As such, we repeated the balancing procedure with 20 random seeds. For each combination of pattern, frame, and seed, we then computed the offset between lost and won cases (\emph{losing} minus \emph{winning}) and measured the significance of such an offset. In the end, we considered the average offset value across seeds for each pattern and frame combination and highlighted the offset as significant if 80\% of the seeds showed a significant offset for that combination. Figure~\ref{fig:usage_before-after_multiseed} summarizes these robustness results. The most relevant findings are robust: communication in won cases showed a higher frequency of decrease in \emph{prognostic} and \emph{motivational} frames after the election, and communication in lost cases was characterized by greater stability in the use of the \emph{motivational} frame.

\begin{figure}[!t]
    \centering   
    \includegraphics[width=0.45\textwidth]{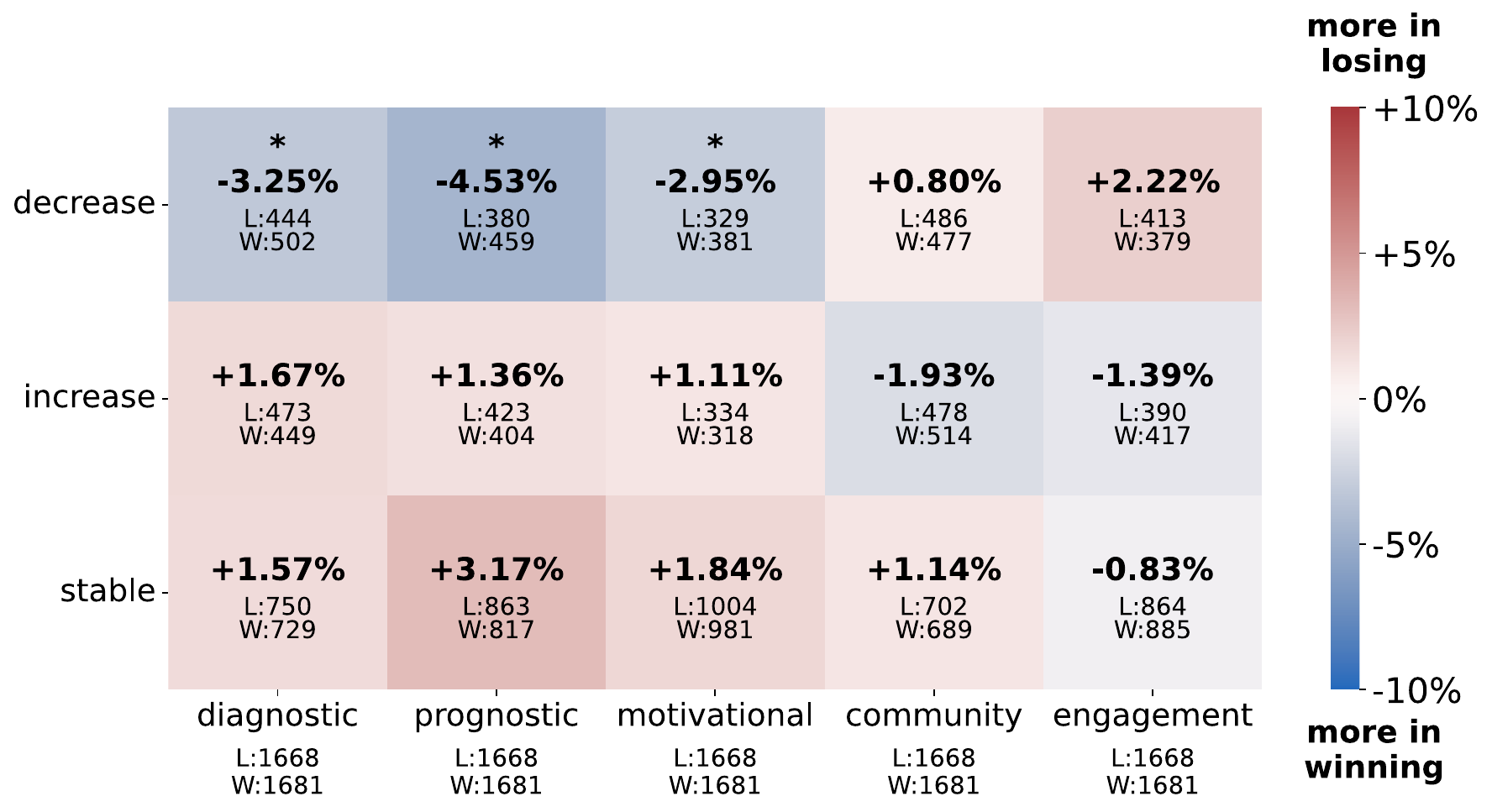}
    \caption{Usage of frames before and after the event, comparing lost and won cases. 
    The $\star$ sign indicates a significant difference between outcome types \rev{($p < 0.05$)}.}
    \label{fig:usage_before-after_robust}
\end{figure}

To account for potential imbalance in the number of representation cases per union, we limited the total cases per union to the 90th percentile, repeating the analyses across 20 random seeds. For each seed and frame, post-/pre-election changes were classified into three patterns: decrease, stable, or increase. 
For each seed, we first balanced the number of wins and losses per union.
Then, for each combination of pattern, frame and seed, we computed the offset between lost and won cases (\emph{losing} minus \emph{winning}) and measured the significance of such an offset. In the end, we considered the average offset value across seeds for each pattern and frame combination and highlighted the offset as significant if 80\% of the seeds showed a significant offset for that combination.
Figure~\ref{fig:usage_before-after_robust} summarizes these robustness results. While not all patterns remain significant under this procedure, the most relevant findings are robust: communication in won cases showed a higher frequency of decrease in \emph{prognostic} and \emph{motivational} frames after the election.

\section{Details on Labor Unions}
\setcounter{figure}{0}
\setcounter{table}{0}
\setcounter{equation}{0}
\renewcommand{\thefigure}{E\arabic{figure}}
\renewcommand{\thetable}{E\arabic{table}}
\renewcommand{\theequation}{E\arabic{equation}}

\begin{figure}[ht!]
    \centering   
    \includegraphics[width=0.38\textwidth]{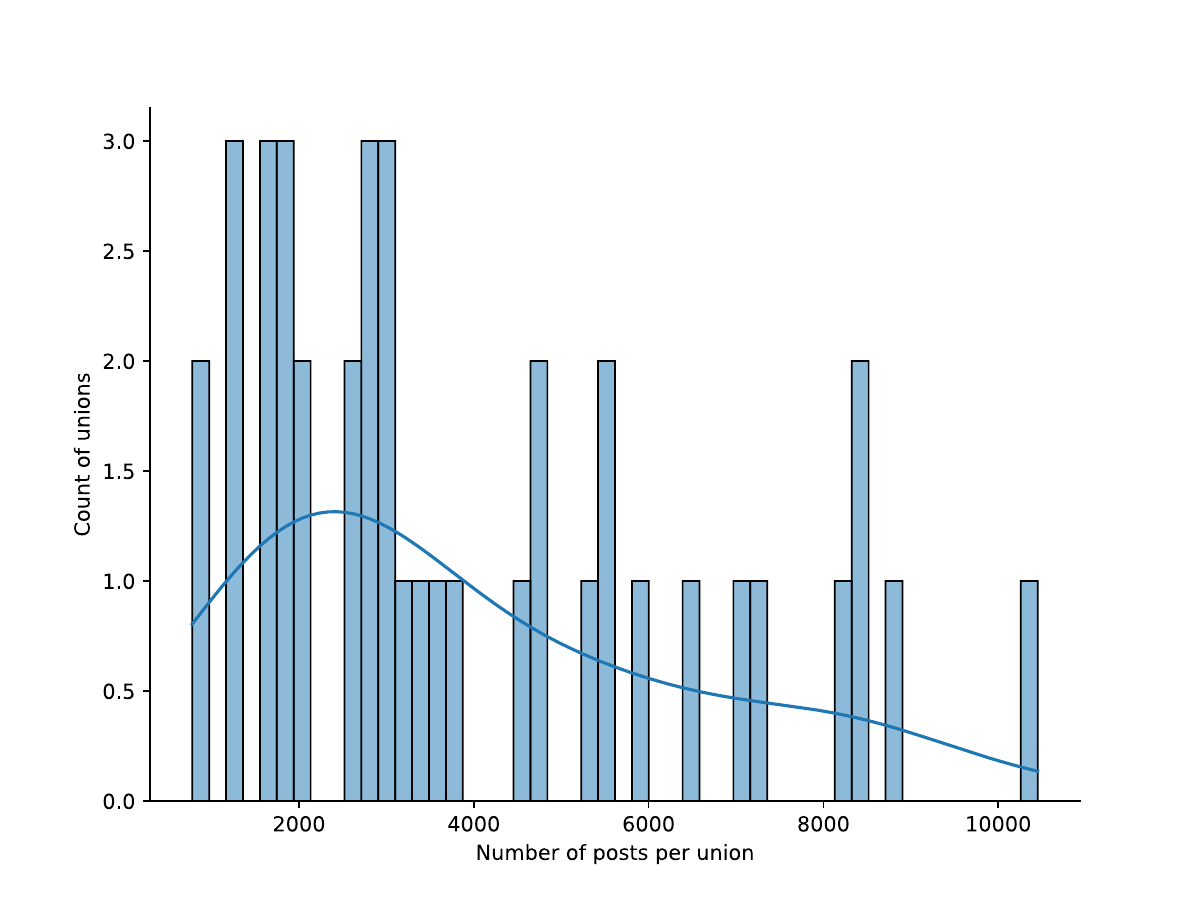}
    \caption{Number of Facebook posts by labor union.}
    \label{fig:n_post_union}
\end{figure}

\rev{Table \ref{tab:union-acronyms} reports the full names unions considered in the study. Figure \ref{fig:n_post_union} shows the distribution of the number of Facebook posts per union.}

\begin{table*}[!t]
\centering
\begin{tabular}{l p{12cm} l}
\hline
\textbf{Acronym} & \textbf{Expanded Name} & \textbf{Union Structure} \\
\hline
RWDSU & Retail, Wholesale and Department Store Union & Industrial\\
UAW & United Automobile Workers & Industrial \\
IBEW & International Brotherhood of Electrical Workers & Craft \\
AFT & American Federation of Teachers & Industrial \\
NATCA & National Air Traffic Controllers Association & Craft \\
IFPTE & International Federation of Professional and Technical Engineers & Industrial \\
AFSCME & American Federation of State, County and Municipal Employees & Industrial \\
NNU & National Nurses United & Industrial \\
IUPAT & International Union of Painters and Allied Trades & Craft \\
Ironworkers & International Association of Bridge, Structural, Ornamental and Reinforcing Iron Workers & Craft \\
NEA & National Education Association & Industrial \\
UMWA & United Mine Workers of America & Industrial \\
AFGE & American Federation of Government Employees & Industrial \\
IATSE & International Alliance of Theatrical Stage Employees & Craft \\
UFCW & United Food and Commercial Workers & Industrial \\
SIU & Seafarers International Union of North America & Industrial \\
SEIU & Service Employees International Union & Industrial \\
SAG-AFTRA & Screen Actors Guild – American Federation of Television and Radio Artists & Craft \\
BCTGM & Bakery, Confectionery, Tobacco Workers and Grain Millers & Industrial \\
APWU & American Postal Workers Union & Industrial \\
UWUA & Utility Workers Union of America & Industrial \\
BAC & Bricklayers and Allied Craftworkers & Craft \\
SMART & Sheet Metal, Air, Rail and Transportation Workers & Craft \\
AFM & American Federation of Musicians & Craft \\
CWA & Communications Workers of America & Industrial \\
USW & United Steelworkers & Industrial \\
UNITE HERE & Unite Here Union & Industrial \\
IAM & International Association of Machinists and Aerospace Workers & Industrial \\
IBT & International Brotherhood of Teamsters & Industrial \\
UA & United Association of Journeymen and Apprentices of the Plumbing and Pipe Fitting Industry & Craft \\
TWU & Transport Workers Union of America & Industrial \\
IAFF & International Association of Fire Fighters & Craft \\
ATU & Amalgamated Transit Union & Industrial \\
IBB & International Brotherhood of Boilermakers, Iron Ship Builders, Blacksmiths, Forgers and Helpers & Craft \\
Roofers & United Union of Roofers, Waterproofers and Allied Workers & Craft \\
LIUNA & Laborers’ International Union of North America & Craft \\
WGAE & Writers Guild of America East & Craft \\
OPEIU & Office and Professional Employees International Union & Industrial \\
CSEA & Civil Service Employees Association & Industrial \\
MEBA & Marine Engineers’ Beneficial Association & Craft \\
\hline
\end{tabular}
\caption{Union acronyms, expanded names, and organizational structure (Craft vs. Industrial).}
\label{tab:union-acronyms}
\end{table*}

\end{document}